\pdfoutput=1
\documentclass[11pt]{article}

\usepackage{graphicx}
\usepackage{times}
\usepackage{url}
\usepackage{multirow}
\usepackage{amsmath}
\usepackage{subfigure}
\usepackage{xcolor}
\usepackage{comment}
\usepackage{url}
\usepackage{booktabs}
\usepackage{cite}
\usepackage{soul}
\usepackage{caption}

\setstcolor{red}

\renewcommand{\paragraph}[1]{\smallskip\noindent {\bf #1}~}

\begin{document}

\title{The Design and Implementation of a Rekeying-aware Encrypted
	Deduplication Storage System}
\author{
Chuan Qin$^1$, Jingwei Li$^2$, Patrick P. C. Lee$^1$\\
$^1$The Chinese University of Hong Kong\\
$^2$Center for Cyber Security, University of Electronic Science and Technology of China\\
Technical Report\\
{\em jwli@uestc.edu.cn, \{cqin, pclee\}@cse.cuhk.edu.hk}
}


\maketitle

\begin{abstract}
Rekeying refers to an operation of replacing an existing key with a new key
for encryption. It renews security protection, so as to protect against key
compromise and enable dynamic access control in cryptographic storage.
However, it is non-trivial to realize efficient rekeying in encrypted
deduplication storage systems, which use deterministic content-derived
encryption keys to allow deduplication on ciphertexts. We design and
implement REED, a rekeying-aware encrypted deduplication storage system.  
REED builds on a deterministic version of all-or-nothing transform (AONT),
such that it enables secure and lightweight rekeying, while preserving the
deduplication capability.  We propose two REED encryption schemes that trade
between performance and security, and extend REED for dynamic access control.
We implement a REED prototype with various performance optimization techniques
and demonstrate how we can exploit similarity to mitigate key generation
overhead.  Our trace-driven testbed evaluation shows that our REED prototype
maintains high performance and storage efficiency.
\end{abstract}

\section{Introduction}

Data explosion has raised a scalability challenge to cloud storage management.
For example, Aberdeen Research \cite{csaplar11} reports that the average size
of backup data for a medium-size enterprise is 285TB, and meanwhile, faces an
annual growth rate of about 24-27\%.  {\em Deduplication} is one plausible
solution that makes storage management scalable. Its idea is to eliminate the
storage of redundant messages that have identical content, by keeping only one
message copy and referring other redundant messages to the copy through
small-size pointers.  Deduplication is shown to effectively reduce
storage space for some workloads, such as backup data \cite{wallace12}.
It has also been deployed in today's commercial cloud storage services (e.g.,
Dropbox, Google Drive, Bitcasa, Mozy, and Memopal) for saving maintenance
costs \cite{harnik10}. 

To protect against content leakage of outsourced data, cloud users often want
to store encrypted data in the cloud.  Traditional symmetric encryption is
incompatible with deduplication: it assumes that users encrypt messages with
their own distinct keys, and hence identical messages of different users will
lead to distinct ciphertexts and prohibit deduplication.  Bellare {\em et al.}
\cite{bellare13a} define a cryptographic primitive called {\em message-locked
encryption (MLE)}, which derives the encryption key from the message itself
through a uniform derivation function, so that the same message
deterministically returns the same ciphertext through symmetric encryption.
One well-known instantiation of MLE is {\em convergent encryption (CE)}
\cite{douceur02}, which uses the cryptographic hash of the message content as
the derivation function.  Storage systems that realize CE or MLE have been
extensively studied and evaluated in the literature (e.g.,
\cite{adya02,cox02,storer08,wilcox-ohearn08,anderson10,bellare13b}).  We
collectively refer to them as {\em encrypted deduplication storage} systems,
which encrypt the stored data while preserving the deduplication capability. 

However, existing encrypted deduplication storage systems do not address 
{\em rekeying}, an operation that replaces an existing key with a new 
key so as to renew security protection.  Rekeying is critical not only for
protecting against key compromise that has been witnessed in real-life
accidents \cite{debian-security-advisory08,kaminsky11,us-computer-emergency-readiness-team14}, but also for enabling dynamic
access control to revoke unauthorized users from accessing data in
cryptographic storage \cite{kallahall02,fu06,backes06,puttaswamy11}.
However, realizing efficient rekeying in encrypted deduplication storage is
challenging.  Since the encryption key of each message in MLE is obtained from
a deterministic derivation function (e.g., a hash function), if we renew the
key by renewing the derivation function, any newly stored message encrypted by
the new key can no longer be deduplicated with the existing identical message;
if we re-encrypt all existing messages with the new key obtained from the
renewed derivation function, there will be tremendous performance overheads
for processing large quantities of messages. 

This paper presents {\em REED}, a \underline{re}keying-aware
\underline{e}ncrypted \underline{d}eduplication storage system that aims for
secure and lightweight rekeying, while preserving identical content for
deduplication.  REED augments MLE with the idea of {\em all-or-nothing
transform (AONT)} \cite{rivest97}, which transforms a secret into a package,
such that the secret cannot be recovered without knowing the entire package.
REED constructs a package based on a deterministic variant of AONT
\cite{li15} and encrypts a small part of the package with a key that is
subject to rekeying, while the remaining large part of the package still
preserves identical content for deduplication.  We show that this approach
enables secure and lightweight rekeying, and simultaneously maintains
deduplication effectiveness.  The contributions of this paper are summarized
as follows.
\begin{itemize}
\item
We propose two encryption schemes for REED, namely basic and enhanced, that
trade between performance and security.  Both schemes enable lightweight
rekeying, while the enhanced scheme is resilient against key leakage through a
more expensive encryption than the basic scheme. 
\item
We extend REED with dynamic access control. We demonstrate how REED integrates
existing primitives, namely ciphertext-policy attribute-based encryption
(CP-ABE) \cite{bethencourt07} and key regression \cite{fu06}, so as to control
the access privileges to different files. 
\item
We exploit the similarity property that is commonly found in backup workloads
\cite{bhagwat09} to mitigate the overhead of MLE key generation, while
preserving deduplication effectiveness. 
\item
We implement a proof-of-concept REED prototype.  Our REED prototype leverages
various performance optimization techniques to mitigate both computational and
I/O overheads.  
\item 
We conduct extensive trace-driven evaluation on our REED prototype in a LAN
testbed.  REED shows lightweight rekeying. It only takes 3.4s to re-encrypt an
8GB file with a new key (in active revocation), and maintains high storage
saving (e.g., higher than 97\%) in real-world datasets.  We also demonstrate
the effectiveness of exploiting similarity in mitigating key generation
overhead. 
\end{itemize}

The source code of our REED prototype is now available for download at the
following website: {\bf http://ansrlab.cse.cuhk.edu.hk/software/reed}.


The remainder of the paper proceeds as follows. 
Section~\ref{sec:background} motivates the need of rekeying for
encrypted deduplication storage. 
Section~\ref{sec:overview} defines our threat model and security goals. 
Section~\ref{sec:design} presents the design of REED.
Section~\ref{sec:similarity} explains how we exploit similarity in REED
to mitigate MLE key generation overhead. 
Section~\ref{sec:implementation} presents the implementation details of REED.
Section~\ref{sec:evaluation} presents our evaluation results. 
Section~\ref{sec:related} reviews related work, and finally,
Section~\ref{sec:conclusions} concludes the paper.

\section{Background and Motivation}
\label{sec:background}

\subsection{Encrypted Deduplication Storage}
\label{subsec:encrypted}

Deduplication exploits content similarity to achieve storage efficiency.  Each
message is identified by a {\em fingerprint}, computed as a cryptographic hash
of the content of the message.  We assume that two messages are identical
(distinct) if their fingerprints are identical (distinct), and that the
fingerprint collision of two distinct messages has a negligible probability in
practice \cite{black06}.  Deduplication stores only one copy of identical
messages, and refers any other identical message to the copy using a
small-size pointer.  In this paper, we focus on 
{\em chunk-level deduplication}, which divides file data into fixed-size or
variable-size chunks, and removes duplicates at the granularity of chunks. 
We use the terms ``messages'' and ``chunks'' interchangeably to refer to the
data units operated by deduplication. 

{\em Message-locked encryption (MLE)} \cite{bellare13a} is a cryptographic
primitive that provides confidentiality guarantees for deduplication storage.
It applies symmetric encryption to encrypt a message with a key called the 
{\em MLE key} that is derived from the message itself, so as to produce a
deterministic ciphertext.  Two identical (distinct) messages will lead to
identical (distinct) ciphertexts, so deduplication remains plausible.  A
special case of MLE is {\em convergent encryption (CE)} \cite{douceur02},
which directly uses the message's fingerprint as the MLE key.  

However, MLE (including CE) is inherently vulnerable to brute-force attacks.
Suppose that a target message is known to be drawn from a finite space. Then
an adversary can sample all messages, derive the MLE key of each message, and
compute the corresponding ciphertexts.  If one of the computed ciphertexts
equals the ciphertext of the target message, then the adversary can deduce the
target message.  Thus, MLE achieves security only for {\em unpredictable}
messages \cite{bellare13a}, meaning that the number of candidate messages is
so large that the adversary cannot feasibly check all messages against the
ciphertexts. 
	
To address the unpredictability assumption, DupLESS \cite{bellare13b} 
implements {\em server-aided MLE}. It uses a dedicated key manager to generate
an MLE key for a message based on two inputs: the message's fingerprint and
a system-wide secret that is independent of the message content.  If the key
manager is secure, then the ciphertexts appear to be encrypted with the keys
that are derived from a random key space.  This provides confidentiality
guarantees even for
predictable messages.  Even if the key manager is compromised, DupLESS still
achieves confidentiality for unpredictable messages. To make MLE key
generation robust, DupLESS introduces two mechanisms.  First, it uses the
oblivious pseudo-random function (OPRF) \cite{goldwasser08} to ``blind'' a
fingerprint to be processed by the key manager, such that the key manager can
return the MLE key without knowing the original fingerprint.  Second, the key
manager rate-limits the key generation requests to protect against online
brute-force attacks. 

In this work, we focus on encrypted deduplication storage based on
server-aided MLE.  Like DupLESS, we deploy a dedicated key manager that is
responsible for MLE key generation, so as to be secure against brute-force
attacks. 

\subsection{Rekeying}
\label{subsec:motivation}

We define {\em rekeying} as the generic process of updating an old
key to a new key in encrypted storage, such that the old key will be revoked,
and all subsequently stored files will be encrypted by the new key.
We argue that rekeying is critical for renewing security protection for
encrypted deduplication storage in two aspects: {\em key protection} and 
{\em access revocation}.  

\paragraph{Key protection:}
There have been real-life cases that indicate how adversaries make key
compromise plausible through various system vulnerabilities, such as design
flaws \cite{debian-security-advisory08,sotirov08,kaminsky11} and programming
errors \cite{us-computer-emergency-readiness-team14}. These threats also apply
to storage systems, since adversaries can compromise file encryption keys and
recover all encrypted files. In addition to key compromise, every
cryptographic key in use is associated with a lifetime, and needs to be
replaced once the key reaches the end of its lifetime \cite{barker12}.
Rekeying is thus critical for key protection.  By immediately updating the
compromised or expired keys, we ensure that the stored files remain protected
by the new keys.   


Since deduplication implies the sharing of data across multiple files and
users, rekeying in encrypted deduplication storage is more critical than
traditional encrypted storage without deduplication. In particular, the
security of a message depends on its MLE key. The leakage of the MLE key may
imply the compromise of multiple files that share the corresponding message. 


\paragraph{Access revocation:} Organizations increasingly outsource
large-scale projects to cloud storage providers for efficient management. We
consider a special case in genome research. 
Genome researchers increasingly leverage cloud services for genome data storage
due to the huge volume of genome datasets \cite{stein10}. Some cloud services,
such as Google Genomics \cite{googlegenomics} and Amazon \cite{amazon14}, have
also set up specific platforms for organizing and analyzing genome
information.  With deduplication, the storage of genome data can be
significantly reduced, for example, by 83\% in real deployment
\cite{NetApp08}.   However, some genome datasets, such as those produced by
disease sequencing projects, are potentially identifiable and must be
protected. 
Thus, dataset owners must properly protect the deduplicated genome data with
encryption and multiple dimensions of access control \cite{NIH15}. When a
researcher leaves a genome project, it is necessary to revoke the researcher's
access privilege to the genome data.


Rekeying can be used to revoke users' access rights by re-encrypting
ciphertexts (e.g., the genome data in the previous example) with new keys and
making old keys inactive.
There are two revocation approaches for existing stored data \cite{backes06}:
(i) {\em lazy revocation}, in which re-encryption of a stored file is deferred
until the next update to the file, and (ii) {\em active revocation}, in which
the stored files are immediately re-encrypted with the new key for up-to-date
protection, at the expense of incurring additional performance overheads.

\subsection{Challenges}
\label{subsec:challenge}

Enabling rekeying in encrypted deduplication storage is a non-trivial issue.
MLE keys are often derived from messages via a global {\em key derivation
function}, such as a hash function in CE \cite{douceur02} or a keyed
pseudo-random function in DupLESS \cite{bellare13b}. A straightforward
rekeying approach is to update the key derivation function directly.  However,
this approach compromises deduplication.  Specifically, a new message cannot
be deduplicated with the existing identical message, because the messages are
now encrypted with different MLE keys that are derived from different
derivation functions.  If we re-encrypt all existing messages with new MLE
keys, the re-encryption overhead will be significant due to the high volume of
stored data. 

There are other possible rekeying approaches, but we argue that they have
limitations.  One approach is based on {\em layered encryption}
\cite{anderson10,rahumed11}. 
Each deduplicated message is first encrypted
with its MLE key, and the MLE key is further encrypted with a master key
associated with each user.  The security now builds on the master key.
Rekeying can simply be done by updating the master key, and re-encrypting the
MLE key with the new master key. This approach does not change the MLE key, so
any new message can be deduplicated with the existing identical message.  Its
drawback is that every ciphertext remains encrypted by the same MLE key.  If
an MLE key is leaked, then the corresponding message can be identified.
Another approach is {\em proxy re-encryption} \cite{ateniese06}, which
transforms a ciphertext encrypted with an old key into another ciphertext
encrypted with a new key.  However, proxy re-encryption is a public-key
primitive and is inefficient when encrypting large-size messages.  

\section{Overview}
\label{sec:overview}

REED is a rekeying-aware encrypted deduplication storage system designed for a
single enterprise or organization in which multiple users want to outsource
storage to a remote third-party cloud provider.  
It deploys a remote server to run deduplication on the storage workloads, and
stores the unique data after deduplication in the cloud provider.
We target the workloads that have high content similarity, such as backup or
genome data (Section~\ref{subsec:motivation}), so that deduplication can
effectively remove duplicates and improve storage efficiency.

REED aims to achieve secure and lightweight rekeying, while preserving
deduplication capability.  In particular, it enables dynamic access control by
controlling which group of users can access a file.  It supports both lazy and
active revocations (Section~\ref{subsec:motivation}); for the latter, the
stored files can be re-encrypted with low overhead.  

\subsection{Architecture}
\label{subsec:arch}

\begin{figure}[!t]
\centering
\includegraphics[width=4in]{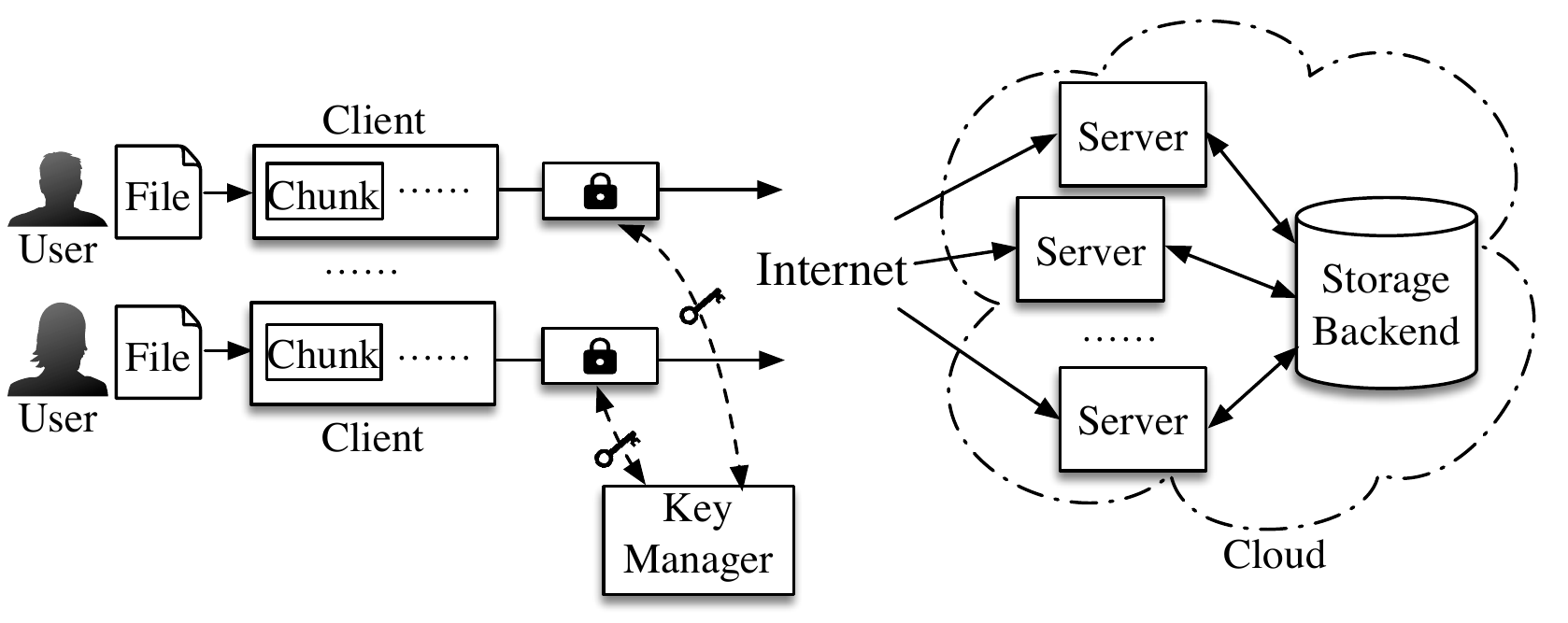}
\caption{REED architecture.}
\label{fig:arch}
\end{figure}

Figure~\ref{fig:arch} presents an overview of the architecture of REED.  REED
follows a client-server architecture. It is composed of different entities, as
described below. 

\paragraph{Client:} 
In each user machine, we deploy a 
{\em REED client} (or {\em client} for short) as a software layer that
provides a secure interface for a user to access and manage files in remote
storage.  To perform an upload operation, the client takes a file
(e.g., a snapshot of a file system folder) as an input from its co-located
user machine.  It divides the file data into chunks, encrypts them, and
uploads the encrypted chunks to the cloud.  We assume that the file has a
sufficiently large size (e.g., GB scale), and can be divided into a large
number of chunks of small sizes (e.g., KB scale).  

We support both fixed-size and variable-size chunking schemes. We implement
variable-size chunking using Rabin fingerprinting \cite{rabin81}, which takes
the minimum, maximum, and average chunk sizes as inputs.  We fix the minimum
and maximum chunk sizes at 2KB and 16KB, respectively, and vary the average
chunk size in our evaluation.  In file downloads, the client reassembles
collected chunks into the original file. 

\paragraph{Key manager:}
As in DupLESS \cite{bellare13b}, REED deploys a {\em key manager} to
provide an interface for a client to access MLE keys for encrypted storage. 
Each client communicates with the key manager to perform necessary
cryptographic operations. We implement server-aided MLE as in DupLESS to  
protect all chunks, including predictable and unpredictable ones, as well as
the OPRF-based MLE key generation protocol (Section~\ref{subsec:encrypted}).  
We elaborate the key generation details in Section~\ref{subsec:keygen}.  Other
approaches, such as blinded BLS signatures \cite{boneh01}, can be used to
implement MLE key generation.  This work considers a single key
manager, while our design can be generalized to multiple key managers for
improved availability \cite{duan14}.  


\paragraph{Server:}
REED performs server-side deduplication. In the cloud, we deploy a 
{\em REED server} (or {\em server} for short) for
storage management.  The server maintains a fingerprint index that keeps track
of all chunks that have been uploaded to the cloud. For a given received
chunk, the server checks by fingerprint if the chunk has already been uploaded
by the same or a different client. If the chunk is new, it stores the chunk
and inserts the chunk fingerprint to the index.  We can deploy multiple
servers for scalability. 

\paragraph{Storage backend:}
Finally, the server stores the encrypted chunks and metadata in the 
{\em storage backend} of the cloud.  For example, if we choose Amazon's cloud
services, we can rent an EC2 virtual machine for a REED server, and use S3 as
the storage backend.

\subsection{Threat Model}
\label{subsec:threat}

We consider an {\em honest-but-curious} adversary that aims to learn the
content of the files in outsourced storage.  The adversary can take the
following actions.  First, it can compromise the cloud (including any hosted
server and the storage backend) to have full access to all stored chunks and
keys.  Also, it can collude with a subset of unauthorized or revoked clients,
and attempt to learn the files that are beyond the access scope of the
colluded clients.  Furthermore, it can monitor the activities of the clients,
identify the MLE keys returned by the key manager, and attempt to extract the
files owned by the monitored clients. 

Our threat model makes the following assumptions. We assume the communication
between a client and the key manager is encrypted and authenticated (e.g.,
using SSL/TLS), so as to defend against any eavesdropping activity in the
network.  Each client and the key manager adopt oblivious key generation
\cite{bellare13b}, so that the key manager cannot infer the fingerprint
information and learn the message content.  We also assume that the key
manager is deployed in a fully protected zone, and an adversary cannot
compromise or gain access to the key manager.  

We do not consider the threat in which an adversary launches online 
brute-force attacks from a compromised client against the key manager, since
the key manager can rate-limit the query rate of each client
\cite{bellare13b}. REED can be deployed in conjunction with remote data
checking \cite{ateniese07,juels07} to efficiently check the integrity of
outsourced files against malicious corruptions.  
REED performs server-side deduplication to protect against the side-channel
attacks mentioned in \cite{harnik10,halevi11}. 

\subsection{Design Goals}
\label{subsec:goals}

Given the threat model, REED focuses on the following design goals.

\begin{itemize}
\item 
{\bf Confidentiality:} REED protects outsourced chunks, such that the chunk
contents are kept secret against any honest-but-curious adversary (e.g., any
unauthorized user or cloud). In addition, REED prevents revoked users from
accessing any new file or update.
\item 
{\bf Integrity:} REED ensures chunk-level integrity of outsourced files.  When
a client downloads a chunk, it can check if the chunk is intact or corrupted. 
\item 
{\bf Practical rekeying:} REED enables rekeying and dynamic access control,
such that it can control which group of users can access a file. It
supports both lazy and active revocations with low overhead; for the latter,
the stored files can be efficiently re-encrypted. REED also allows an
unlimited number of rekeying operations.  
\item 
{\bf High storage efficiency:}  REED achieves storage efficiency by
deduplication. In addition, it introduces small storage overhead due to keys
or metadata.  
\item
{\bf High encryption performance:}  REED introduces limited encryption
overhead when compared to the network transmission via the cloud. 
\end{itemize}

\section{REED Design}
\label{sec:design}

We now present the design details of REED.  We propose two encryption schemes
that trade between performance and security. We also demonstrate how REED
realizes dynamic access control using existing primitives.  Finally, we
analyze the security of REED.  In this section, we focus on the essential
features of REED that support secure and lightweight rekeying, without
considering workload characteristics; in Section~\ref{sec:similarity}, we
exploit similarity for performance improvements. 

\subsection{Main Idea}
\label{subsec:idea}

REED builds security simultaneously on two types of symmetric keys: a
file-level secret key per file (or {\em file key} for short) and a chunk-level
MLE key for each chunk (or {\em MLE key} for short).  During rekeying, REED
only needs to renew the file key, while the MLE keys of all chunks remain
unchanged.  

REED uses \emph{all-or-nothing transform} (AONT) \cite{rivest97} as the
underlying cryptographic primitive. AONT is an unkeyed, randomized encryption
mode that transforms a message into a ciphertext called the {\em package},
which has the property that it is computationally infeasible to be reverted
back to the original message without knowing the entire package.  
The original AONT design prohibits deduplication, since its transformation
takes a random key as an input to construct a package. Thus, REED uses 
{\em convergent AONT (CAONT)} \cite{li15}, which replaces the random key with
a deterministic message-derived key to construct a package. This ensures that
identical messages always lead to the same package. 

REED augments CAONT to enable rekeying.  Our insight is to achieve security by
sacrificing a slight degradation of storage efficiency.  The idea of REED is
based on AONT-based secure deletion \cite{peterson05}, which makes the entire
package unrecoverable by securely removing a small part of a package.  REED
extends the idea to make it applicable for rekeying.  Specifically, REED
generates a CAONT package with the MLE key as an input, and encrypts a small
part of the package, called the {\em stub} \cite{peterson05}, with the file
key.  Thus, the entire package is now protected by both the file key and the
MLE key.  The stub size is small; for example, our implementation sets it as
64~bytes, equivalent to 0.78\% for an 8KB chunk.  In addition, we can still
apply deduplication to the remaining large part of the package, called the
{\em trimmed package}, so as to maintain storage efficiency.

In the following, we first design two rekeying-aware encryption schemes on a
per-chunk basis (Section~\ref{subsec:encryption}), followed by enabling REED
with dynamic access control on a per-file basis. 



\subsection{Encryption Schemes}
\label{subsec:encryption}

We propose the basic and enhanced encryption schemes for REED.  The basic
scheme is more efficient, but is vulnerable to the leakage of an MLE key.  On
the other hand, the enhanced scheme protects against the leakage of an MLE
key, while introducing an additional encryption step.  In the following, we
first explain the basics of AONT \cite{rivest97} and its variant CAONT
\cite{li15}, followed by how the basic and enhanced encryption schemes build
on CAONT.

\paragraph{All-or-nothing transform (AONT):} AONT \cite{rivest97} works as
follows. It transforms a message $M$ to a package denoted by $(C, t)$, where
$C$ and $t$ are called the \emph{head} and \emph{tail}, respectively.
Specifically, it first selects a random encryption key $K$ and generates a
pseudo-random mask  ${\sf G}(K) = {\sf E}(K, S)$, where ${\sf E}(\cdot)$
denotes a symmetric key encryption function (e.g., AES-256) and $S$ is a
publicly known block with the same size as $M$.  It then computes $C = M
\oplus {\sf G}(K)$, where `$\oplus$' is the XOR operator, and also computes $t
= {\sf H}(C) \oplus K$, where ${\sf H}(\cdot)$ is the hash function (e.g.,
SHA-256).   Note that the resulting package has a larger size than the
original message $M$ by the size of $t$.  To recover the original message $M$,
suppose that the whole package $(C,t)$ is known.  We first compute 
$K = {\sf H}(C) \oplus t$, followed by computing $M = C \oplus {\sf E}(K, S)$. 


CAONT \cite{li15} follows the same paradigm of AONT, but replaces the random
encryption key $K$ by a deterministic cryptographic hash $h = {\sf H}(M)$
derived from the message $M$.  This ensures that packages generated by
identical messages remain identical, and hence the packages can still be
deduplicated.  Another feature of CAONT is that it allows integrity checking
without padding.  Specifically, after the package is reverted, the integrity
can be verified by computing the hash value of $M$ and checking if it equals
$h$.

\begin{figure}[t]
\centering
\includegraphics[scale=.65]{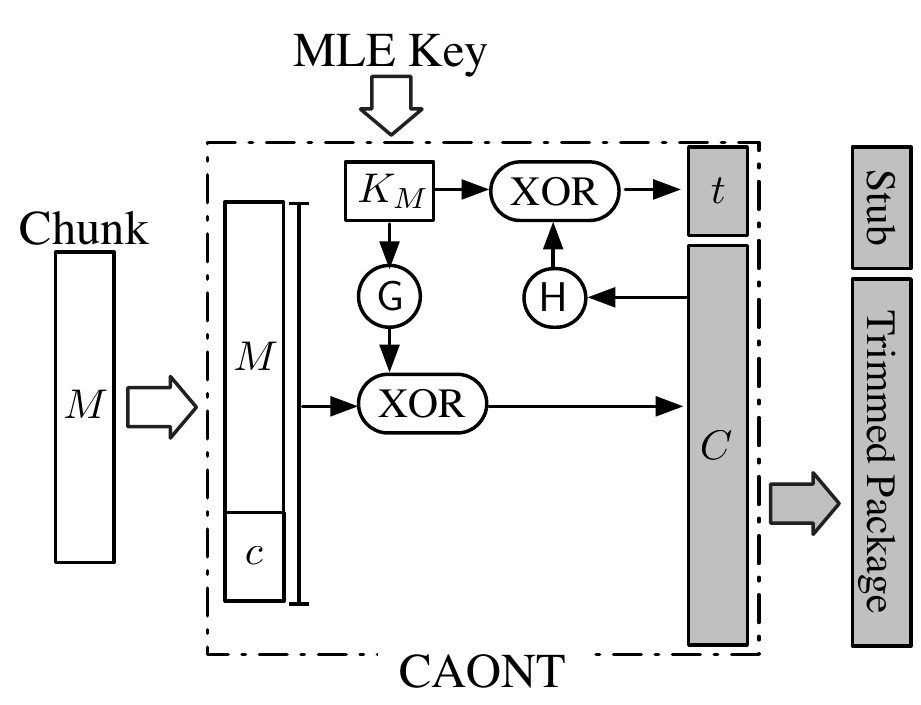}
\caption{Basic encryption of REED.}
\label{fig:basic}
\end{figure}

\paragraph{Basic encryption:} 
The basic encryption scheme leverages CAONT
\cite{li15} to generate both the trimmed package and the stub, as shown in
Figure~\ref{fig:basic}. In particular, we make two modifications to CAONT.
The first modification is to replace the cryptographic hash key in CAONT
\cite{li15} by the corresponding MLE key $K_M$ generated by the key manager.
The rationale is that we use the MLE key to achieve security even for
predictable chunks through server-aided MLE \cite{bellare13b} 
(Section~\ref{subsec:encrypted}).  However, we now cannot use the hash key for
integrity checking as in CAONT.  Thus, the second modification is to append a
publicly known, fixed-size canary $c$ to $M$ \cite{resch11} for CAONT, so that
the integrity of $M$ can be checked.  In our implementation, we set the
fixed-size canary $c$ to be 32~bytes of zeroes. 

The basic encryption scheme is detailed as follows. We first
concatenate an input chunk $M$ with the canary $c$ to form $(M||c)$, and
compute the pseudo-random mask ${\sf G}(K_M) = {\sf E}(K_M, S)$, where $K_M$
is the MLE key obtained from the key manager  and $S$ is the publicly known
block with the same size of $(M||c)$.  We compute the package head $C = (M||c)
\oplus {\sf G}(K_M)$, and the package tail $t = K_M \oplus {\sf H}(C)$.  We
generate the stub by trimming the last few bytes (e.g., 64~bytes) from the
package $(C,t)$, and leave the remaining part as the trimmed package.
Finally, we encrypt the stub with the file key.  Reconstruction of a message
works reversely, and we omit details here. 

\begin{figure}[t]
\centering
\includegraphics[scale=.65]{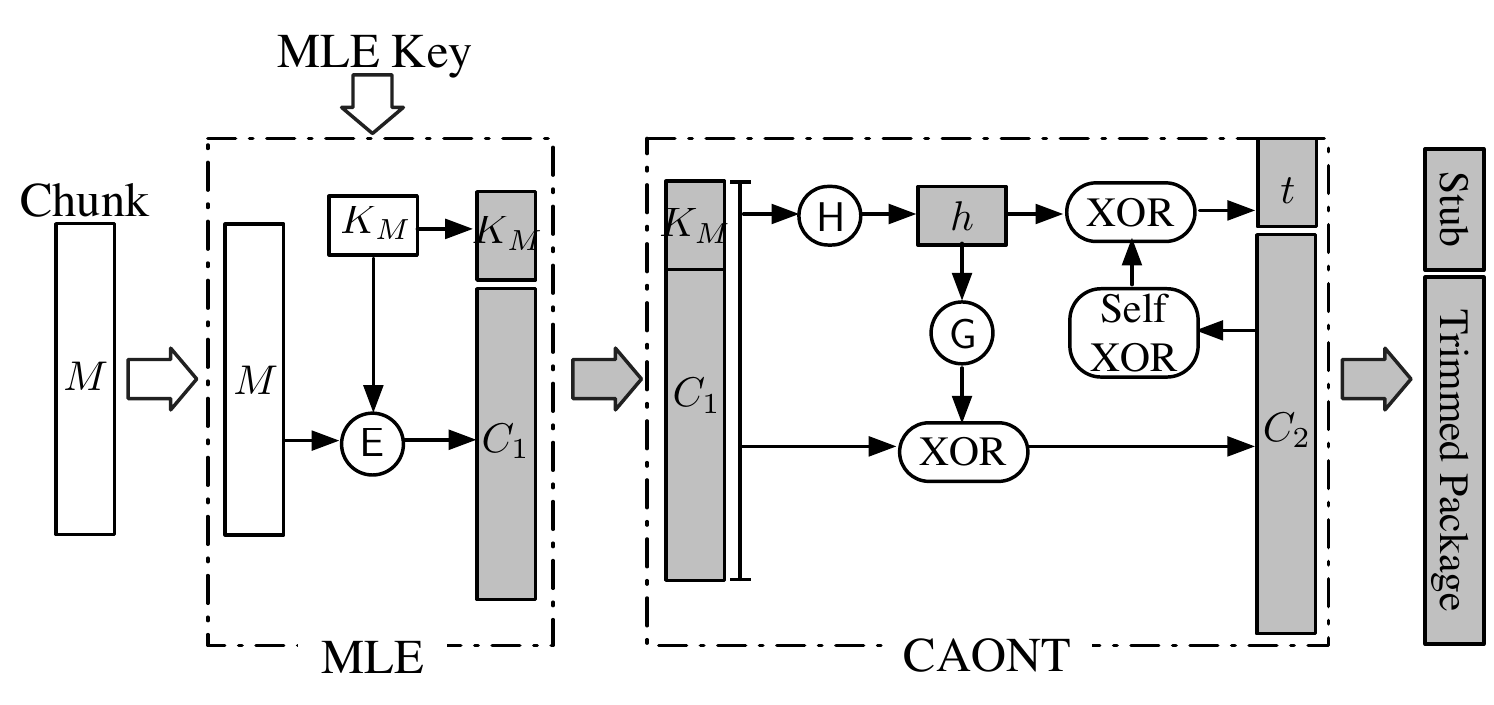}
\caption{Enhanced encryption of REED.}
\label{fig:enhanced}
\end{figure}

We briefly comment on the security guarantees of the basic encryption scheme.
The security of each chunk builds on both the file key and the MLE key.  If
both the file key and the MLE key are secure, then given both the trimmed
package and the encrypted stub of a chunk, it is computationally infeasible to
revert them to the original chunk.  In addition, if the file key is renewed,
it is also computationally infeasible to restore the stub (which is now
protected by the new file key) and hence the original chunk using the old file
key.

One limitation of the basic encryption scheme is that it is vulnerable to the
compromise of the MLE key.  Specifically, an adversary can monitor the MLE
keys generated by the key manager at a compromised client
(Section~\ref{subsec:goals}).  If an MLE key is revealed, the adversary can
recover the pseudo-random mask and XOR the mask with the trimmed package to
extract a majority part of the chunk.  



\paragraph{Enhanced encryption:}  
We propose the enhanced encryption scheme, which protects against the
compromise of its MLE key.  Figure~\ref{fig:enhanced} shows the workflow of
the enhanced encryption, which first applies MLE to form a ciphertext,
followed by applying CAONT \cite{li15} to the MLE ciphertext.  The rationale
is that even if an adversary obtains the MLE key, it still cannot recover
original chunk because the MLE ciphertext is now protected by CAONT.  

The enhanced encryption scheme is detailed as follows. First, we encrypt an
input chunk $M$ with the MLE key $K_M$ as in traditional MLE, and obtain the
ciphertext $C_1$.  We then transform the concatenation $C_1 || K_M$ based on the
original CAONT \cite{li15}. We can now use the hash key $h = {\sf H}(C_1 ||
K_M)$, instead of the MLE key used in the basic encryption scheme, 
to transform the
package.  This eliminates the security dependence on the MLE key.  Formally,
we compute the hash key $h = {\sf H}(C_1 || K_M)$ and the pseudo-random mask
${\sf G}(h) = {\sf E}(h, S)$, where $S$ is a publicly known block with the
same size as $C_1 || K_M$, and computes the package head $C_2 = (C_1 || K_M)
\oplus {\sf G}(h)$.  

Since the hash key $h$ allows integrity checking \cite{li15}, we can generate
the tail $t$ with a {\em self-XOR} operation for efficiency \cite{peterson05},
instead of using the cryptographic hash as in the basic encryption scheme
(Figure~\ref{fig:basic}).  Specifically, we evenly divide $C_2$ into a set of
fixed-size pieces, each with the same size as $h$.  We then XOR all the
pieces as well as $h$ to compute the tail $t$. Note that the self-XOR result
cannot be predicted without knowing the entire content of $C_2$.  Finally, we
obtain the trimmed package and the stub from $(C_2, t)$. 

To reconstruct $M$, we first reconstruct $(C_2, t)$ from the trimmed package
and the stub.  We evenly divide $C_2$ into fixed-size pieces, each with the
same size as $t$, and compute $h$ by XOR-ing the pieces and $t$.  We then
recover $C_1||K_M = C_2 \oplus {\sf G}(h)$, and check the integrity by
comparing ${\sf H}(C_1||K_M)$ and $h$.  We finally compute $M = {\sf D}(K_M,
C_1)$, where ${\sf D}(\cdot)$ is the decryption function. 

We now briefly comment on the security guarantees of the enhanced encryption
scheme.  As in basic encryption, the enhanced encryption scheme ensures that
each chunk remains secure if both the file key and the MLE key are secure.  If
the MLE key is leaked, the adversary can recover the original chunk from the
MLE ciphertext (i.e., the input to CAONT), yet the original chunk remains
secure if the unpredictability assumption still holds (see
Section~\ref{subsec:encrypted}).  We present a more detailed security analysis
in Section~\ref{subsec:security}.



\subsection{Dynamic Access Control}
\label{subsec:policy}

\begin{figure}[t]
\centering
\includegraphics[width=2.9in]{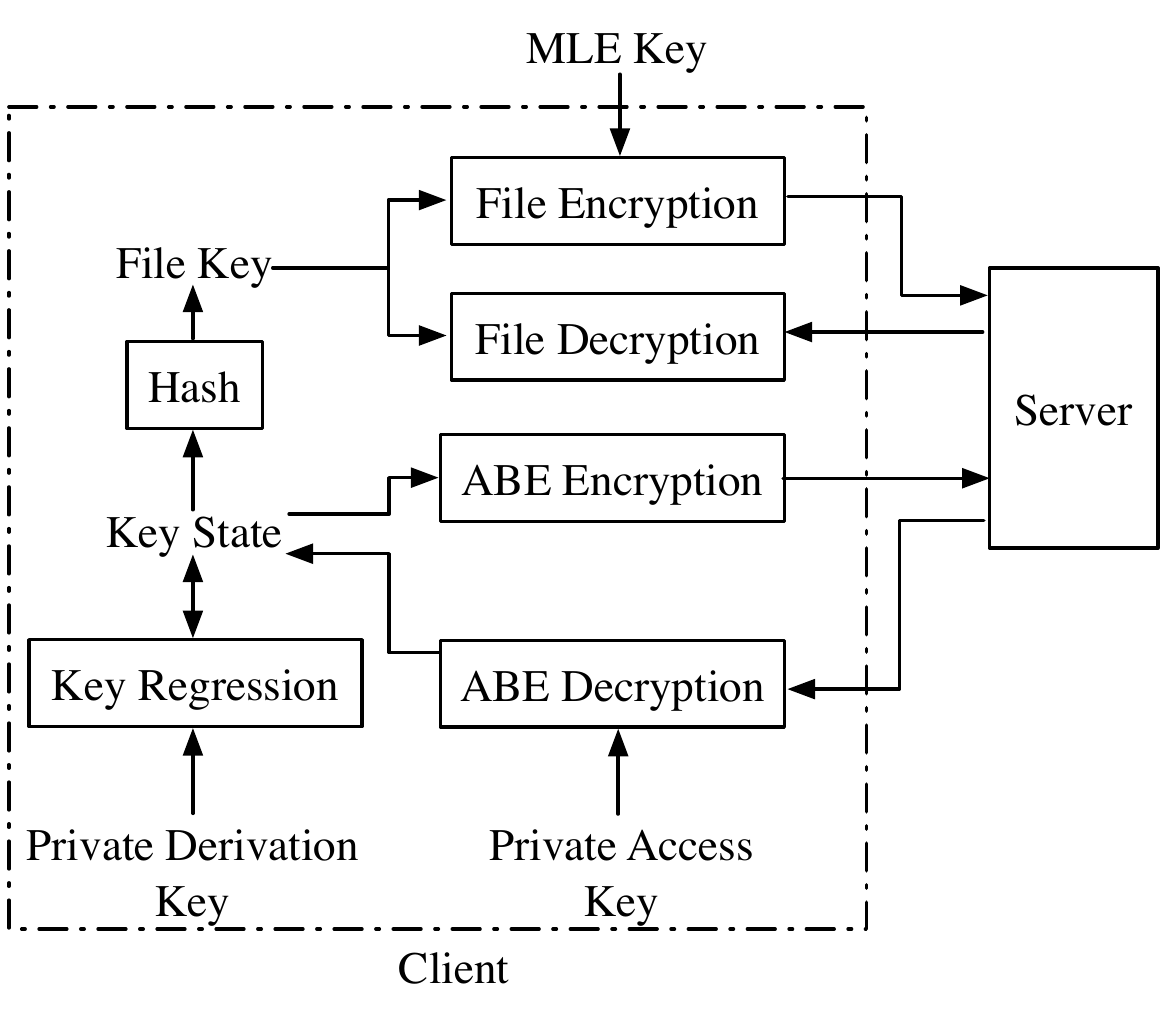}
\caption{REED generates a file key from the hash of a key state.  The key
state is derived from key regression. Its access is protected by CP-ABE.}
\label{fig:keyflow}
\end{figure}

REED supports dynamic access control by associating each
file with a {\em policy}, which provides a specification of which users are
authorized or revoked to access the file.  Our policy-based design builds on
two well-known cryptographic primitives: {\em ciphertext policy
attribute-based encryption (CP-ABE)} \cite{bethencourt07} and 
{\em key regression} \cite{fu06}.  REED integrates both primitives to generate
the corresponding file key, as shown in Figure~\ref{fig:keyflow}.  Note that
our goal here is {\em not} to propose new designs for CP-ABE and key
regression; instead, we demonstrate how REED can work seamlessly with them to
provide advanced security functionalities for rekeying.  In the following, we
elaborate how REED integrates the two primitives. 

\paragraph{Access control:}  REED defines policies based on CP-ABE
\cite{bethencourt07}.  In CP-ABE, a message is encrypted based on a specific
policy that describes which users can decrypt the message.
Each policy is represented in the form of an {\em access tree}, in which each
non-leaf node represents a Boolean gate (e.g., AND or OR), while each leaf node
represents an {\em attribute} that defines or classifies some user property
(e.g., the department that a user belongs to, the employee rank, the contract
duration, etc.).  Each user is given a private key that corresponds to a set of
attributes.  If a user's attributes satisfy the access tree, his private key can
decrypt the ciphertext.  

Our current design of REED treats each attribute as a unique identifier for
each user.  We issue each user with a CP-ABE private key, called the 
{\em private access key}, related to the identifier.  We define the policy of
each file as an access tree that connects the identifiers of all authorized
users with an OR gate.  Thus, any authorized user can decrypt the ciphertext,
which we use to protect the file key (see the rekeying discussion below).  
Note that we can define more attributes and a more sophisticated access
tree structure for better access control.

\paragraph{Rekeying:}  REED supports both lazy and active revocations for
rekeying.  In lazy revocation, REED builds on key regression \cite{fu06},
which is a serial key derivation scheme for generating different versions of
keys.  Specifically, key regression introduces a sequence of {\em key states},
such that the current key state can derive the previous key states, but it
cannot derive any future key state.  Thus, an authorized user can access all
previous key states, and the corresponding files,  by using only the current key
state; meanwhile, a user revoked from the current key state cannot access any
new file that is protected by a future key state.  
REED implements lazy revocation using the RSA-based key regression scheme
\cite{fu06}.  We assign each user with a unique pair of public-private keys
called the {\em derivation keys}, such that the private derivation key is used
to generate new key states for the files owned by the user, while the public
derivation key is used to derive the previous key states.  The file key will be
obtained by generating a cryptographic hash of the current key state.  Each
key state refers to a policy, and it will be encrypted by CP-ABE associated
with the authorized users.  In other words, any authorized user can retrieve
the current key state, and hence the file key, with his private access key. 

REED implements active revocation following the same paradigm as in lazy
revocation, except that the files affected by active revocation are
immediately re-encrypted with the new file key. 

\subsection{Operations}
\label{subsec:operations}

We now summarize the interactions among a client, a server, the key manager,
and the storage backend in REED operations.  We focus on three basic
operations, including upload, download, and rekeying.  

\paragraph{Upload:} To upload a file $F$, the client first picks a random
key state $S_F$ and hashes it into a symmetric file key $\kappa_F$.  It splits
$F$ into a set of chunks $\{M\}$, computes their fingerprints, and runs the
OPRF protocol \cite{bellare13b} with the key manager to obtain the MLE keys
$\{K_M\}$ of these chunks (Section~\ref{subsec:arch}).  For each $M$, it uses
$K_M$ to transform a chunk into a trimmed package and a stub, using either the
basic or enhanced encryption scheme (Section~\ref{subsec:encryption}).  
The client writes
the stubs of all the chunks of the same file into a separate {\em stub file}
for storage, and the stub file will be encrypted by the file key $\kappa_F$.  
In addition, the client generates a
{\em file recipe}, which includes the file information such as the file
pathname, file size, and the total number of chunks.  Furthermore, the client
encrypts $S_F$ using CP-ABE based on the policy of the file.  Finally, the
client uploads the following information to the REED server: (i) the trimmed
packages and encrypted stubs for all chunks, (ii) file recipe, and (iii) the
encrypted key state $S_F$ and the metadata that includes the policy
information.  Note that we do not need to upload MLE keys, as they are not
used in decryption (Section~\ref{subsec:encryption}).  The server performs
deduplication on the received trimmed packages.  All information will be
stored at the storage backend. 


\paragraph{Download:} To download a file $F$, the client first retrieves the
encrypted key state $S_F$ and decrypts it with the private access key.  It then
hashes $S_F$ to recover the file key $\kappa_F$.  In addition, it downloads
all trimmed packages and encrypted stubs from the storage backend, with the
help of the REED server and the file recipe.  It decrypts the stubs via
$\kappa_F$, and finally reconstructs all chunks for $F$.  Note that if the
client detects any tampered chunk, the reconstruction operation will abort.

\paragraph{Rekeying:} To rekey $F$ with new access privileges, the client
(on behalf of the owner of $F$) retrieves $S_F$ and its metadata, and decrypts
$S_F$ with the private access key. It then generates a new key state $S_F'$
based on key regression (Section~\ref{subsec:policy}).  It encrypts $S_F'$
via CP-ABE based on a new policy (e.g., with a new group of users).  It finally
uploads the encrypted $S_F'$ as well as its metadata that describes the new
policy information.  For active revocation, the client also downloads the
stubs of $F$, re-encrypts them with a new file key obtained by hashing $S_F'$,
and finally uploads the re-encrypted stubs. 

\subsection{Security Analysis}
\label{subsec:security}

We now analyze the security of REED based on our security goals. 

\paragraph{Confidentiality:} 
We show how REED achieves confidentiality at three levels. First, an adversary
can access all trimmed packages, encrypted stubs, and encrypted key states
from a compromised server.  Since the adversary cannot compromise any private
access key and private derivation key, all trimmed packages and encrypted
stubs cannot be reverted.  Thus, REED achieves the same level of
confidentiality like DupLESS \cite{bellare13b}
(Section~\ref{subsec:encrypted}). 

Second, an adversary can collude with revoked or unauthorized clients, through
which the adversary can learn a set of private derivation keys and private
access keys.  Due to the protection of CP-ABE and key regression, these
compromised private keys cannot be used to decrypt the file key ciphertexts
beyond their access scopes. Without proper file keys, the adversary cannot
infer anything about the underlying chunks. One special note is that a client
may keep the MLE key (in basic encryption) or the hash key (in enhanced
encryption) of a chunk in CAONT (Figures~\ref{fig:basic} and
\ref{fig:enhanced}, respectively) to make the chunk accessible even after
being revoked.  However, if the chunk is updated, the revoked client cannot
learn any information from the updated chunk because CAONT will use a new MLE
key or hash key to transform the updated chunk, making the old one useless. 
	  

Finally, an adversary can monitor a subset of clients and identify the MLE
keys requested by them. The enhanced encryption scheme of REED ensures
confidentiality for unpredictable chunks, even though the victim clients are
authorized to access these chunks.  Specifically, the enhanced encryption
scheme builds an additional security layer with the file key. 
As long as the file key is secure, it is computationally infeasible to restore
the MLE ciphertext (i.e., the input to CAONT) due to the protection of CAONT.
Note that the adversary may restore the MLE ciphertext by launching a
brute-force attack to check if the MLE ciphertext is transformed into the
trimmed package through CAONT, but it is computationally infeasible if chunks
are unpredictable (see Section~\ref{subsec:encryption}).  Thus, identifying an
MLE key does not help recover the original chunk, and hence the original chunk
remains secure. 


\paragraph{Integrity:}   Both the basic and enhanced encryption schemes of
REED ensure chunk-level integrity, such that any modification of the trimmed
package or the stub of a chunk can be detected. 
In the basic encryption scheme, the MLE key can be reverted as $K_M = {\sf
H}(C) \oplus t$ (Section~\ref{subsec:encryption}).  Since ${\sf H}(C)$
depends on every bit of $C$ \cite{webster85}, the modification of any part of
the package will lead to an incorrect $K_M$. Thus, the client can easily
detect the modification by checking the canary padded with the reverted chunk.

Using similar reasonings, the enhanced encryption scheme also ensures the
integrity of a chunk, such that a client performs integrity checking by
comparing if ${\sf H}(C_1 || K_M)$ equals $h$
(Section~\ref{subsec:encryption}).  One special note regarding the enhanced
scheme is that its use of the self-XOR operation may return a correct hash key
$h$ even if the package is tampered.  For example, an intelligent adversary
can divide $C_2$ into fixed-size pieces and flip the same bit position for an
even number of the pieces.  On the other hand, a tampered package will be
reverted to a wrong input even with the correct hash key, and its integrity
violation can be caught by comparing it with $h$.

\subsection{Discussion}
\label{subsec:discussion}

We present some open issues of our current REED design. 

\paragraph{Fault tolerance:}
In this work, we do not explicitly address fault tolerance. To improve fault
tolerance of stored data, we can distribute both trimmed packages and stubs
across multiple cloud providers via deduplication-aware secret sharing
\cite{li15}.


\paragraph{Metadata management:}
We currently focus on the encryption and rekeying for file chunks, while we do
not address those for file metadata (e.g., file recipe). We can obfuscate
sensitive metadata information, such as the file pathname, by encoding it via
a salted hash function. 

\paragraph{Group-based file management:}
We currently perform rekeying on a per-file basis. We can generalize rekeying
for file group with multiple files. This makes file management more flexible.
On the other hand, we need to define new metadata to describe the file group
information.

\section{Exploiting Similarity}
\label{sec:similarity} 

REED builds on server-aided MLE key generation \cite{bellare13b}, in which the
key manager generates an MLE key for each message (or chunk in our case).  In
this section, we argue that MLE key generation is expensive and significantly
degrades the overall performance of REED.   In view of this, we propose to
exploit the similarity feature that is commonly found in backup workloads, so
as to mitigate the performance overhead of REED. 

\subsection{Overhead of MLE Key Generation}
\label{subsec:keygen}

Recall that REED realizes the OPRF protocol to ``blind'' MLE key generation
as in DupLESS \cite{bellare13b}.  In our design, we configure the key manager
with a system-wide public/private key pair, based on 1024-bit RSA in our case.
Let $e$ and $d$ be the public and private keys, respectively, and $N$ be the
modulus.  For each chunk to be uploaded, a client performs MLE key generation
in the following steps (note that all arithmetic is performed in modulo $N$).
\begin{itemize}
\item
{\em Blind}: the client selects a random number $r$, raises it to power $e$,
and multiplies $r^e$ with the fingerprint. It sends the blinded fingerprint to
the key manager.
\item
{\em Sign}: the key manager computes an RSA signature by raising the blinded
fingerprint to power $d$. It returns the result to the client.  Note that the
key manager does not know the original fingerprint, which is ``blinded'' by
the random number $r$. 
\item
{\em Unblind}: the client multiplies the received result with the inverse of
$r$.  It also hashes the unblinded result to form the MLE key. 
\end{itemize}

OPRF-based MLE key generation is expensive, especially when it operates on
small-size chunks.  Its overhead comes from two aspects.  First, if a client
sends individual per-chunk MLE key generation requests to the key manager,
there will be substantial transmission overhead.  Also, since the OPRF
protocol for key generation is based on public key cryptography, there will be
substantial computational overhead due to modular exponentiation. 

Table~\ref{tab:breakdown} provides a performance breakdown (in terms of
latency) of MLE key generation for an 8KB chunk.  We obtain average results
over 10 runs from our experimental testbed (Section~\ref{sec:evaluation}). If
we implement all the steps serially, the total latency for MLE key generation
is 1125.3$\mu$s, or equivalently the throughput is only
$\frac{\rm{8KB}}{\rm{1125.3}\mu\rm{s}}\approx$~6.9MB/s.  If we deploy REED in
a Gigabit LAN (our experimental testbed), MLE key generation easily becomes a
performance bottleneck of REED.  In particular, the sign operation occupies
48\% of the total latency, and it cannot be trivially parallelized for
performance improvement. 

\begin{table}[!t]
\caption{Performance breakdown of MLE key generation for an 8KB chunk.}
\label{tab:breakdown}
\renewcommand{\arraystretch}{1.1}
\vspace{-3pt}
\centering
\begin{small}
\begin{tabular}{|l|c|}
\hline
\multicolumn{1}{|c|}{\bf Steps} & {\bf Latency ($\mu$s)} \\
\hline
\hline
Blind (performed by the client) & 46.3 \\
\hline
Sign (performed by the key manager) & 537.2 \\
\hline
Unblind (performed by the client) &  246.9 \\
\hline
\hline
Round-trip transmission &  294.9 \\
\hline
\end{tabular}
\end{small}
\end{table}

\subsection{Limitations of Simple Optimizations}
\label{subsec:simple}

To mitigate MLE key generation overhead, our conference paper \cite{li16} uses
two optimization approaches: (i) {\em batching} per-chunk MLE key generation
requests and (ii) {\em caching} the most recently generated MLE keys in the
client's local key cache.  While both approaches can mitigate key generation
overhead based on evaluation results, they still have the following
limitations.  

Batching per-chunk MLE key generation requests aims to reduce round-trip
transmission overhead, but it does not reduce computational overhead (see
Table~\ref{tab:breakdown}).  As shown in our conference paper \cite{li16},
batching 256 per-chunk key generation requests for 8KB chunks can only
achieve a key generation speed of 17.64MB/s, which is still much smaller than
the network speed in a Gigabit LAN. 

Caching the MLE keys is effective in mitigating key generation overhead, based
on the observation that the adjacent uploads of a client often share high
content similarity; for example, backup snapshots for a file system are highly
similar if there are only small changes to the file system.  As shown in our
conference paper \cite{li16}, we can eliminate most key generation requests
for the uploads of subsequent backups after the first one, so the upload
speeds for subsequent backups are almost network-bound (around 100MB/s).
However, the caching approach has few limitations.  First, it is only
effective for uploads that are largely duplicated with the previous one (e.g.,
it is ineffective for the first backup \cite{li16}).  Second, its required
local cache space is not scalable; for example, it needs 4GB of cache space
per 1TB of storage, assuming that we configure 8KB chunks and 256-bit MLE
keys.  Finally, it is unreliable due to the volatile nature of cache. 

\subsection{Similarity-based Approach}
\label{subsec:similarity}

We propose a similarity-based approach for MLE key generation, such that we
can mitigate MLE key generation overhead, while preserving deduplication
effectiveness.  First, we adopt coarse-grained MLE key generation on a larger
data unit called {\em segment}, which comprises multiple adjacent chunks and
has a size on the order of megabytes (e.g., 1MB by default in our case).  To
form a segment, we implement the variable-size segmentation scheme in
\cite{lillibridge09} that operates directly on chunk fingerprints and is
configured by the minimum, average, and maximum segment sizes.  Specifically,
we traverse the stream of chunks, and place a segment boundary after the chunk
if the chunk fingerprint modulo a pre-defined {\em divisor} is equal to a
fixed constant (which we set to -1 as in \cite{lillibridge09}).  Here, the
divisor is configured by the average segment size to specify the expected
number of chunks between adjacent segment boundaries.
We ensure that the segment size is at least the minimum segment size,
and we always place a boundary after the chunk whose inclusion makes the
segment size larger than the maximum segment size.  In our implementation, we
vary the average segment size, and fix the minimum segment size and maximum
segment size as half and double of the average segment size, respectively.  

Clearly, per-segment MLE key generation incurs much fewer key generation
requests than the per-chunk one, thereby significantly mitigating the overall
performance overhead.  On the other hand, segment-level MLE key generation can
introduce different segment-level MLE keys for different segments (and hence
ciphertexts), even though the segments share a large portion of identical
chunks.  This compromises deduplication effectiveness. 

Thus, our similarity-based approach aims to maximize deduplication
effectiveness by carefully generating segment-level MLE keys. Our insight is
to assign ``similar'' segments with the same MLE key and encrypt every chunk
of a segment with the corresponding segment-level MLE key.  If two ``similar''
segments share a large number of identical chunks, the identical chunks are
still encrypted with the same key and hence deduplicated. 

In this work, we borrow the Extreme Binning approach \cite{bhagwat09} to
identify similar segments.  Specifically, for each segment that contains
multiple chunks, a client selects the chunk (called the {\em representative
chunk}) whose fingerprint value is the minimum.  It uses the minimum
fingerprint to request the key manager for the segment-level MLE key.  It
then encrypts each chunk (via either basic or enhanced encryption of
REED) with the received MLE key.  The rationale is that if two segments share
a large number of identical chunks, there is a high probability that both
segments share the same representative chunk (due to Border's Theorem
\cite{broder97}). 

\begin{figure}[t]
\centering 
\includegraphics[scale=0.8]{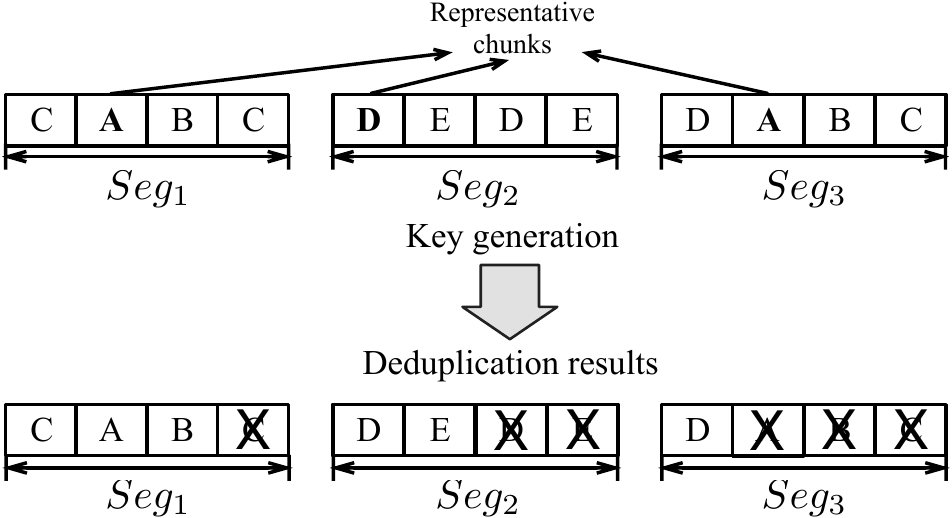}
\caption{Example of similarity-based MLE key generation.}
\label{fig:key-gen} 
\end{figure}

Figure~\ref{fig:key-gen} shows an example of our similarity-based approach. 
Consider three segments $Seg_{1}, Seg_{2}$ and $Seg_{3}$ that have four chunks
each, and suppose that their representative chunks are $A$, $D$, and $A$,
respectively.  Since both segments $Seg_{1}$ and $Seg_{3}$ share the same
MLE key, their identical chunks (i.e., 
$A$, $B$ and $C$) in these similar segments can be deduplicated.  Note that
the approach cannot achieve exact deduplication; for example, chunk~$D$ in
segments $Seg_{2}$ and $Seg_{3}$ cannot be deduplicated due to the different
segment-level MLE keys.  Nevertheless, since similarity is common in backup
workloads \cite{bhagwat09}, we expect that our similarity-based approach
achieves high deduplication effectiveness, as also validated in our evaluation
(Section~\ref{sec:evaluation}). 

\subsection{Security Analysis}
\label{subsec:impacts}

We now analyze the security impact of the similarity-based MLE key generation
on both the basic and enhanced encryption schemes. Unfortunately, our
similarity-based key generation cannot preserve the confidentiality of chunks
in the basic scheme. The reason is that it uses a segment-level MLE key
(derived from the minimum fingerprint of a segment) as an input to CAONT to
transform all chunks in a segment. This creates the same pseudo-random mask
${\sf G}(.)$ for all chunks in the same segment.  This allows an adversary to
apply an XOR operation to any two of the resulting trimmed packages to remove
the mask, and learn a majority part of the XOR result of original chunks.

Nevertheless, we emphasize that the similarity-based MLE key generation does
not introduce new
security risks in the enhanced scheme. The reason is that the pseudo-random
mask is generated from both the MLE key and the MLE ciphertext (i.e., $C_1$ in
Figure~\ref{fig:enhanced}).  As a result, different chunks lead to different
pseudo-random masks, which are infeasible to be removed without the
knowledge of the file key.  Although the similarity-based MLE key generation
allows an adversary to narrow down the attack space of the online brute-force
attack by requesting the MLE keys for potential minimum fingerprints, the key
manager can lower the rate limit for key generation requests
\cite{bellare13b}.  Since segment-level key generation has already reduced the
number of key generation requests, lowering the rate limit has no impact on
normal users.  Thus, the enhanced encryption scheme can benefit from
similarity-based MLE key generation for performance gains, and achieve similar
performance to the basic encryption scheme based on our evaluation (see
Section~\ref{subsec:synthetic}).

\subsection{Summary}

We summarize the benefits of our similarity-based MLE key generation over the
simple optimizations in Section~\ref{subsec:simple}.  First, it operates on a
per-segment basis, it inherently reduces the number of MLE key generation
requests, independent of the amount of duplicates in the workloads.  Also, it
does not need to locally cache MLE keys, and hence it eliminates the concerns
of maintaining a large cache space.  Finally, it exploits similarity to remove
duplicate chunks to maintain deduplication effectiveness.



\section{Implementation}
\label{sec:implementation}

\begin{figure}[t]
\centering
\includegraphics[width=.8\textwidth]{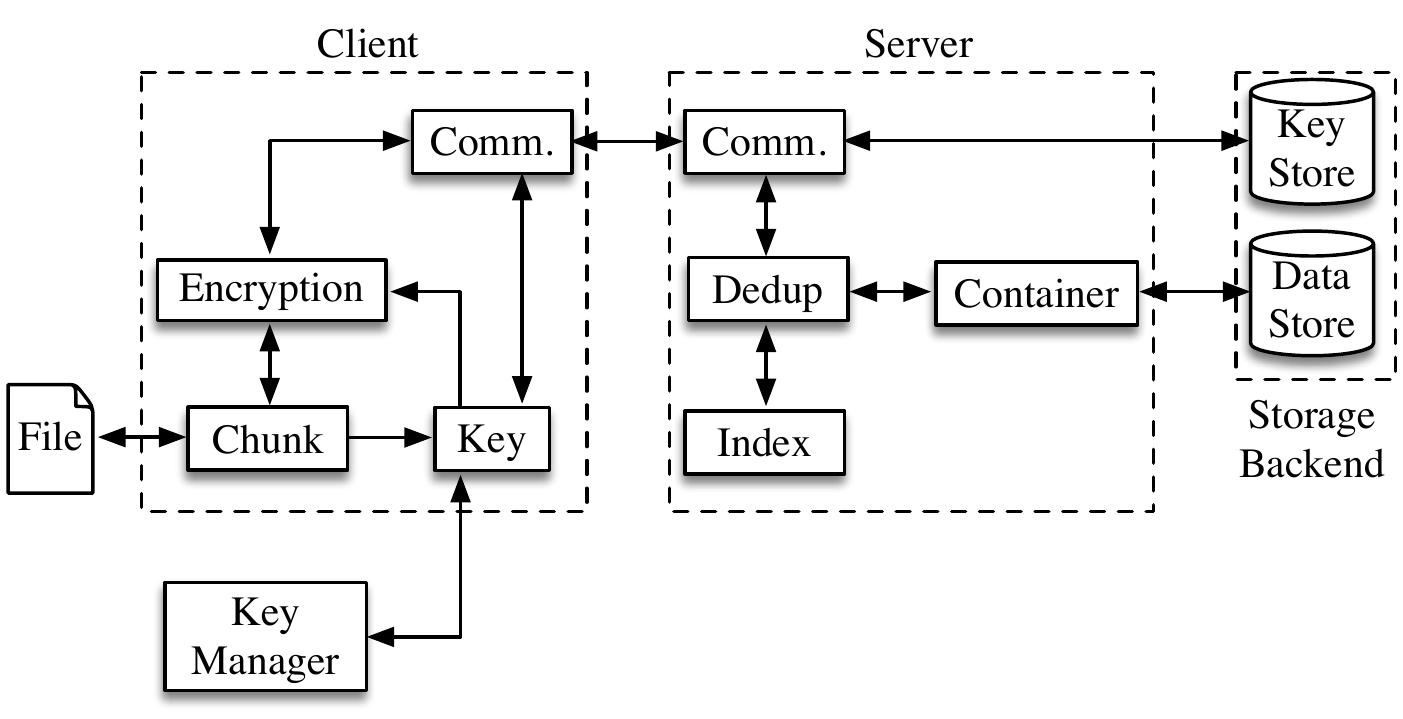}
\caption{Implementation of the REED prototype.}
\label{fig:module}
\end{figure}

We implement a REED prototype in C++ based on our previously built system
CDStore \cite{li15}.   We follow the modular approach as in CDStore to
implement REED, and Figure~\ref{fig:module} shows how the modules of REED are
organized.  We mainly extend CDStore to support rekeying, with the
addition of a key manager, the basic and enhanced encryption schemes
(Section~\ref{subsec:encryption}), dynamic access control
(Section~\ref{subsec:policy}), and the similarity-based key generation
approach (Section~\ref{subsec:similarity}).  We also use OpenSSL 1.0.2a
\cite{openssl} and CP-ABE toolkit 0.11 \cite{bethencourt_abetoolkit} to
implement the cryptographic operations in REED.  The current REED prototype,
including the original CDStore modules, contains around 11,000~LOC.

\paragraph{Client:}
Like CDStore, a client divides an input file into fixed-size or variable-size
(via Rabin fingerprinting \cite{rabin81}) chunks in the \emph{chunk} module.
It can also reassemble collected chunks into the original file during file
download.  We currently use SHA-256 to compute chunk fingerprints.  

In the \emph{key} module, the client runs the OPRF protocol with the key
manager for generating either chunk-based or segment-level MLE keys (see
Section~\ref{subsec:similarity}).  In addition, it implements RSA-based
key regression \cite{fu06} for generating new key states during rekeying, and
protects each key state using CP-ABE \cite{bethencourt07} (via the CP-ABE
toolkit \cite{bethencourt_abetoolkit}). 

In the \emph{encryption} module, the client implements both basic and enhanced
encryption schemes (and the corresponding decryption schemes). In both
encryption schemes, the client transforms a chunk into a trimmed package and a
stub through CAONT, in which we implement ${\sf G}(.)$ via AES-256 and the
hash function ${\sf H}(.)$ via SHA-256 (see Section~\ref{subsec:encryption}). 
To resist brute-force attacks on the stub yet preserving storage
efficiency, we configure the stub size as 64 bytes for each chunk. To enable
integrity checking on reconstructed chunks, we set the fixed-size canary $c$
in both schemes to be 32 and zero bytes. The client encrypts each stub file
(that consists of stubs of the same file) with a file key hashed from the
corresponding key state via SHA-256. 

The \emph{communication} module is similar to that in CDStore. In this module,
the client uploads (resp. downloads) all stored data to (resp. from) the
server, including the trimmed packages, the encrypted stub file, the file
metadata, the encrypted key state, and the public derivation key. 

\paragraph{Key manager:}
A key manager authenticates clients' connections via SSL/TLS. It implements
the OPRF protocol based on 1024-bit RSA, and computes an RSA signature on each
incoming blinded fingerprint.

\paragraph{Server:}
A server can receive file data from multiple clients via the
\emph{communication} module.  It performs deduplication on the trimmed
packages via the \emph{dedup} module, and only stores unique trimmed packages
in the storage backend. Since a file may have a large number of trimmed
packages, the server packs them in units of containers to make storage and
retrieval efficient via the \emph{container} module. Like CDStore, we cap the
container size at 4MB by default.  

In the \emph{index} module, the server keeps track of indexing information,
including the fingerprints of all trimmed packages for deduplication, and the
references to all trimmed packages and file recipes in the storage backend for
file retrieval. 


\paragraph{Storage backend:}
We separate the storage into file data and key information
for better management.  Specifically, we create two stores at the
storage backend: (i) the data store, which stores the file data such as file
recipes, trimmed packages, stub files, and all related file metadata, and (ii)
the key store, which stores the key information such as encrypted key
states. Separating the storage management of key information and file data
gives flexibility, for example, by leveraging a more robust platform for
encryption key management \cite{Cloudencryption13}.

\paragraph{Optimization:} To achieve reasonable performance, REED batches I/O
requests, and also parallelizes the encryption (resp. decryption) operations
of uploaded (resp. downloaded) chunks via multi-threading.  Here, we only
configure two threads for encryption/decryption, as our evaluation results
indicate that two threads are sufficient for achieving the required
performance.

\section{Evaluation}
\label{sec:evaluation}

We evaluate REED on a LAN testbed composed of multiple machines, each of which
is equipped with a quad-core 3.4GHz Intel Core i5-3570, 7200RPM SATA hard
disk, and 8GB RAM, and installed with 64-bit Ubuntu 12.04.2~LTS.  All
machines are connected via a 1Gb/s switch.  

Our default setting of REED is as follows.  We run one REED client, one key
manager, and five REED servers in different machines.  We use multiple REED
servers for improved scalability.  In particular, four of the five servers
manage the data store, and the remaining one server manages the key store. In
practice, both the data store and the key store should be deployed in a shared
storage backend (e.g., cloud storage); however, to remove the I/O overhead of
accessing the shared storage backend in our evaluation, we simply have each
server store information in its local hard disk.  In addition to the default
setting, we describe additional specific settings in each experiment, and also
consider the case where multiple clients are involved.  We compile our
programs with g++ 4.8.1 with the -O3 option.  For performance tests, we
present the average results over 10 runs. We do not include the variance
results in our plots, as they are generally very small in our evaluation.  In
the following, we use a synthetic dataset and two real-world datasets for our
evaluation.  

\subsection{Synthetic Data}
\label{subsec:synthetic}

We evaluate different REED operations through synthetic data.  In particular,
we evaluate how segment-level MLE key generation mitigates overhead
(Section~\ref{sec:similarity}).  Specifically, we generate a 2GB file of
synthetic data with globally unique chunks (i.e., the chunks have no duplicate
content).  Before each experiment, we load the synthetic data into memory to
avoid generating any disk I/O overhead.

\paragraph{Experiment A.1 (MLE key generation performance):}  
We first measure the performance of MLE key generation between a client and
the key manager.  The client creates chunks of the
input 2GB file using variable-size chunking based on Rabin fingerprinting
with a specified average chunk size.  We also group the chunks into
variable-size segments with a specified average segment size
(Section~\ref{sec:similarity}).  The client computes the minimum fingerprint
of each segment and requests for segment-level MLE keys from the key manager.
We measure the {\em MLE key generation speed}, defined as the ratio of the
file size (i.e., 2GB) to the total time starting from when the client creates
the input file until it obtains all segment-level MLE keys from the key
manager. 

\begin{figure}[t]
\centering
\subfigure[Varying chunk size (segment size fixed at 1MB)]{
\includegraphics[width=2.5in]{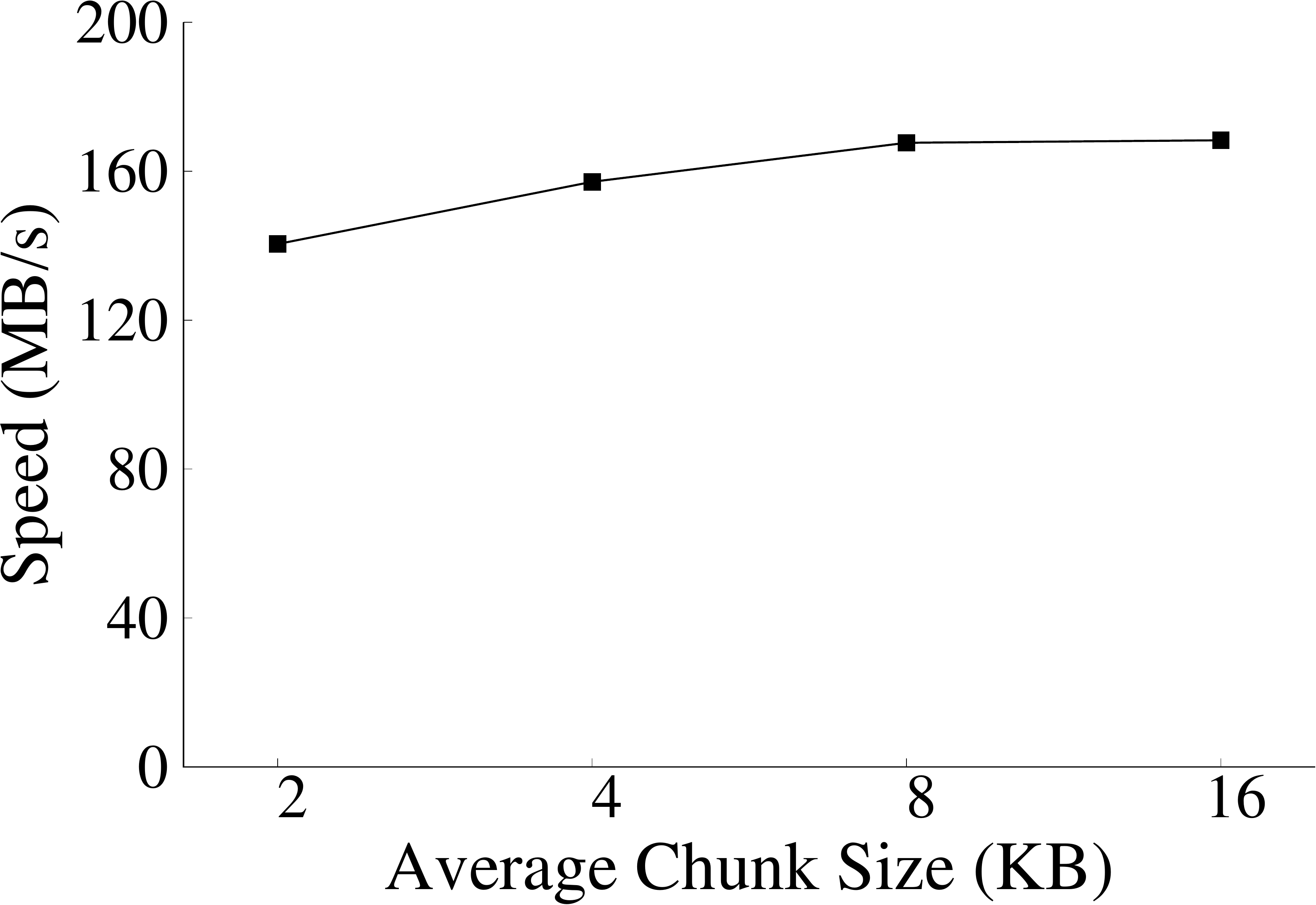}
\label{fig:keygen:chunk}
} 
\hspace{0.2in}
\subfigure[Varying segment size (chunk size fixed at 8KB)]{
\includegraphics[width=2.5in]{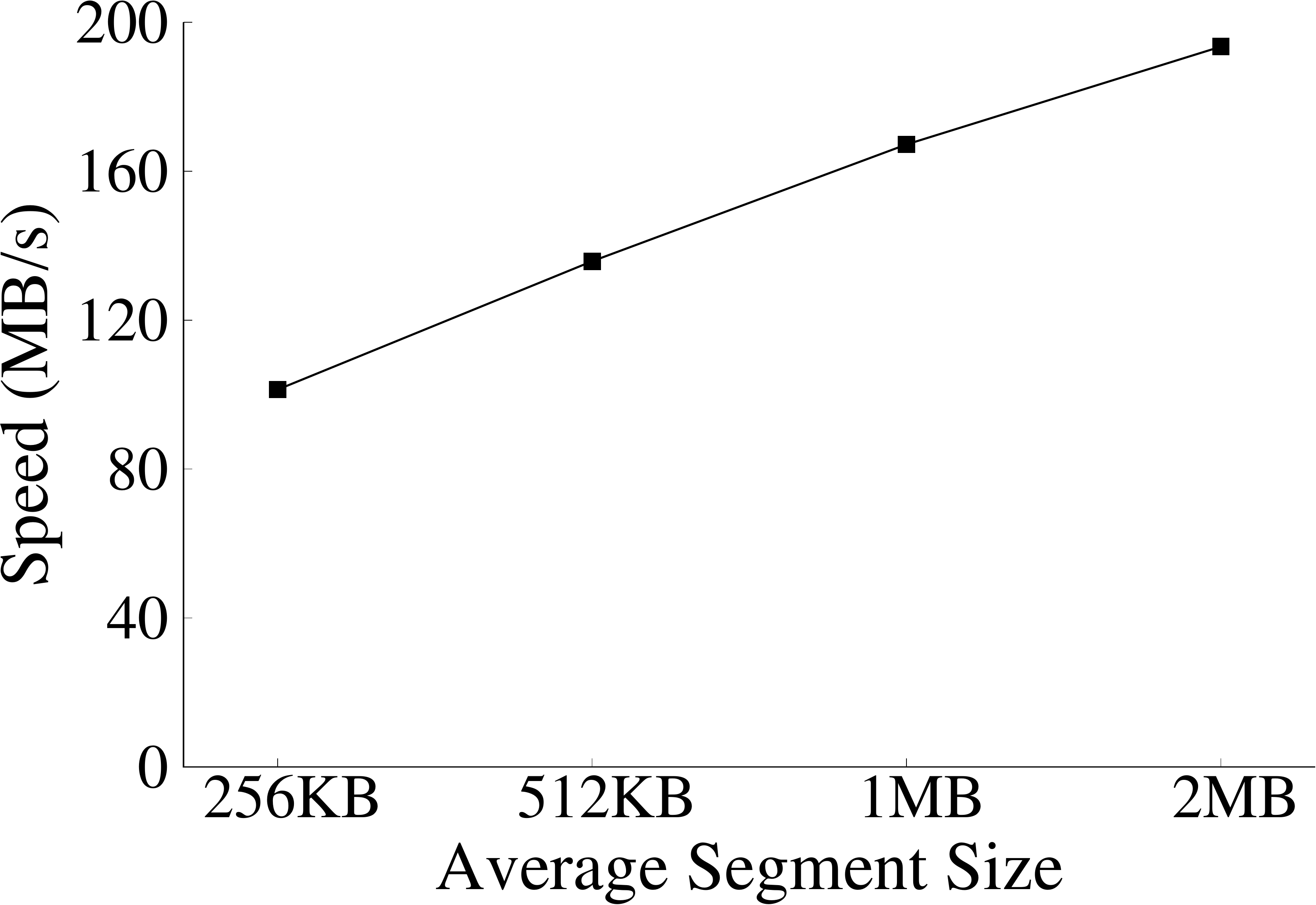}
\label{fig:keygen:segment}
}
\caption{Experiment A.1 (MLE key generation performance).}
\end{figure}

Figure~\ref{fig:keygen:chunk} shows the MLE key generation speed versus
the average chunk size, in which we fix the average segment size as 1MB. We
observe that the speed increases with the average chunk size, mainly because
we process fewer chunks to find the minimum fingerprint for each segment.
When the average chunk size is at least 8KB, the key generation speed becomes
steady at around 168MB/s, since the key manager is now saturated by
segment-level key generation requests and the speed is bounded by the
computation of the key manager. For comparison, our conference paper
\cite{li16} shows a {\em per-chunk} key generation speed is below 20MB/s.

Figure~\ref{fig:keygen:segment} shows the MLE key generation speed versus the
average segment size, in which we fix the average chunk size as 8KB.  The
speed increases with the average segment size, as a larger segment size
implies fewer MLE keys to be generated.  When the segment size is at least
512KB, the key generation speed is above 130MB/s, which is higher than the
network speed in our LAN testbed (i.e., 1Gb/s). 

\paragraph{Experiment A.2 (Encryption performance):}  We measure the
performance of both basic and enhanced encryption schemes.  Suppose that the
client has created chunks with variable-size chunking and obtained MLE keys
from the key manager.  Here, we measure the {\em encryption speed}, defined as
the ratio of the file size (i.e., 2GB) to the total time of encrypting all
chunks into trimmed packages and stubs.  

\begin{figure}
\centering
\includegraphics[width=2.5in]{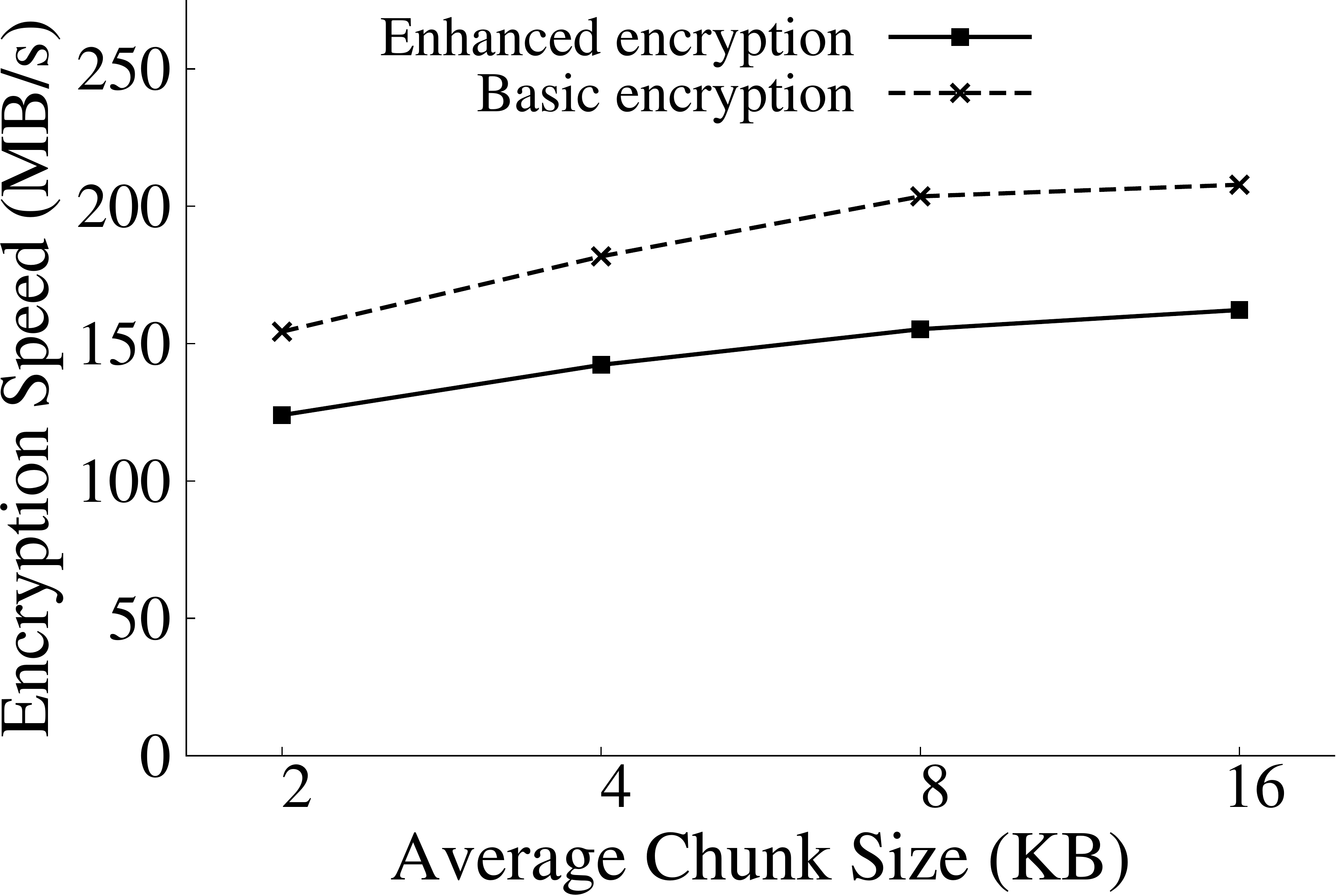}
\caption{Experiment~A.2 (Encryption performance).}
\label{fig:speed:encryption}
\end{figure}

Figure~\ref{fig:speed:encryption} shows the speeds of both basic and enhanced
encryption schemes versus the average chunk size (note that the average segment size has no impact on the
encryption speed).  
The throughput of both encryption schemes increases with the average chunk
size, mainly because fewer chunks need to be processed.  The basic scheme is
faster than the enhanced scheme, as the enhanced scheme introduces an
additional encryption (see Section~\ref{subsec:encryption}).  For example, for
the average chunk size 8KB, the basic scheme has 203MB/s, 24\% faster than
155MB/s in the enhanced scheme.  We observe that the encryption speeds of both
schemes are higher than the network speed (i.e., 1Gb/s), and hence the
encryption speed is not the performance bottleneck in REED.  We further
justify this claim in Experiment~A.3. 

\paragraph{Experiment A.3 (Upload and download performance):}  
We now measure the upload and download performance of REED.  
For performance comparisons, we also include the basic encryption scheme,
although it is shown to be insecure in similarity-based MLE key generation
(see Section~\ref{subsec:impacts}). 
We first consider
the case of a single client.  The client first uploads a 2GB file of unique
data, followed by downloading the 2GB file.  We measure the {\em upload speed}
as the ratio of the file size to the total time of sending all file data to
the servers (including the chunking, key generation, encryption, and data
transfer), and the {\em download speed} as the ratio of the file size to the
total time starting from when the client issues a download request until all
original data is recovered.  

\begin{figure}[t]
\centering
\subfigure[Upload speed versus average chunk size]{
	\includegraphics[width=2.5in]{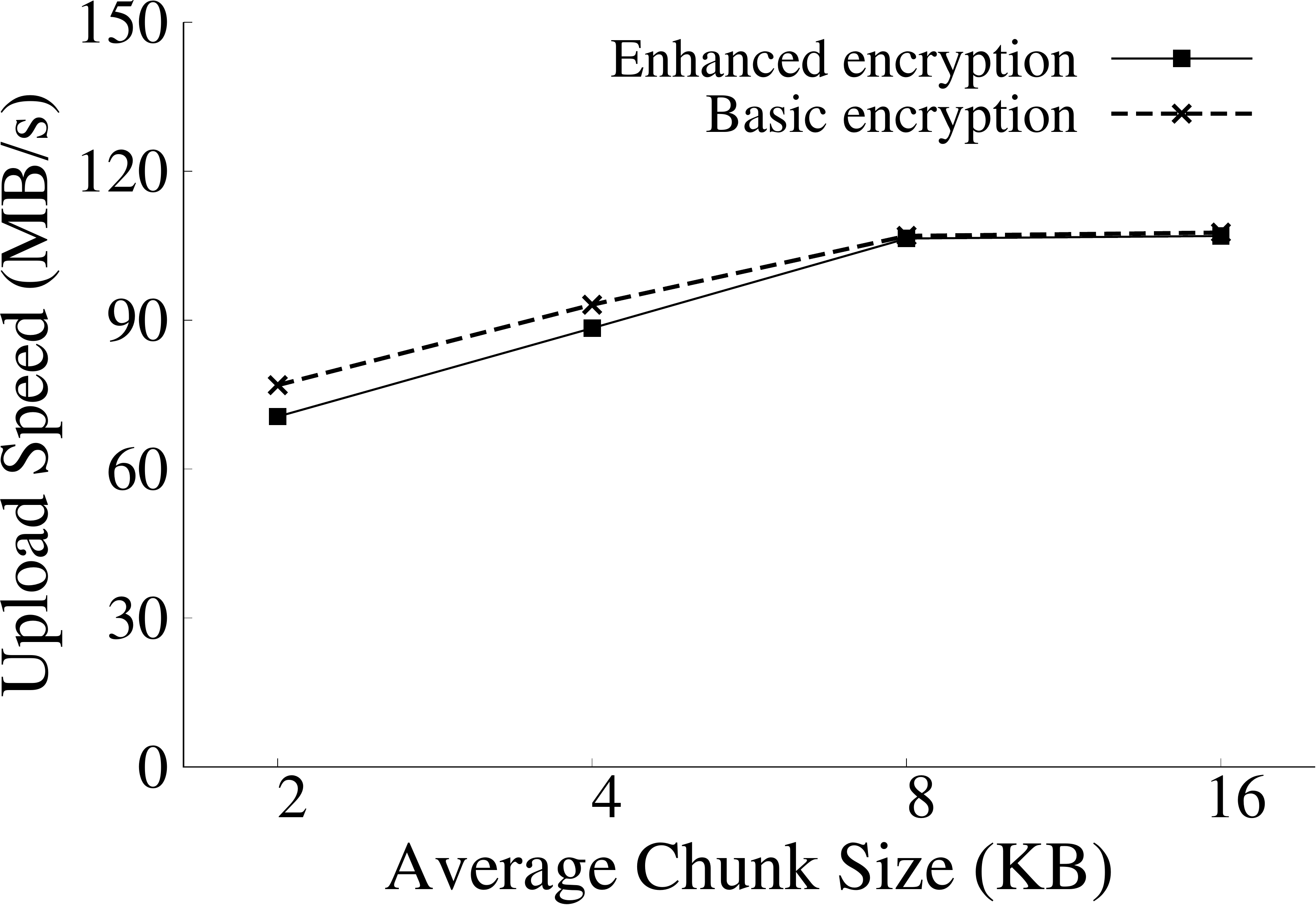}
	\label{fig:speed:up:chunk}
} \hspace{0.2in}
\subfigure[Upload speed versus average segment size]{
	\includegraphics[width=2.5in]{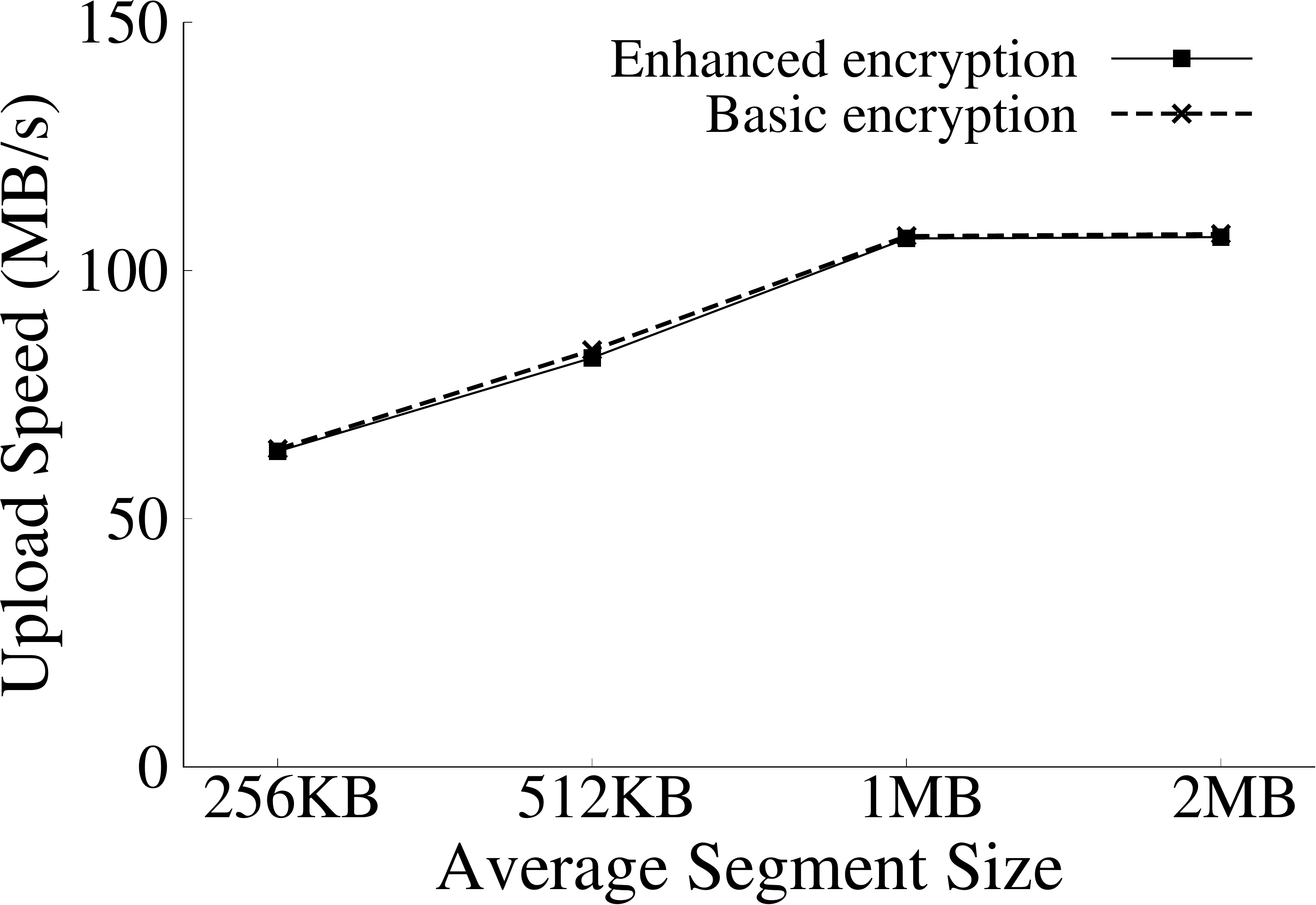}
	\label{fig:speed:up:segment}
} 
\subfigure[Download speed]{
	\includegraphics[width=2.5in]{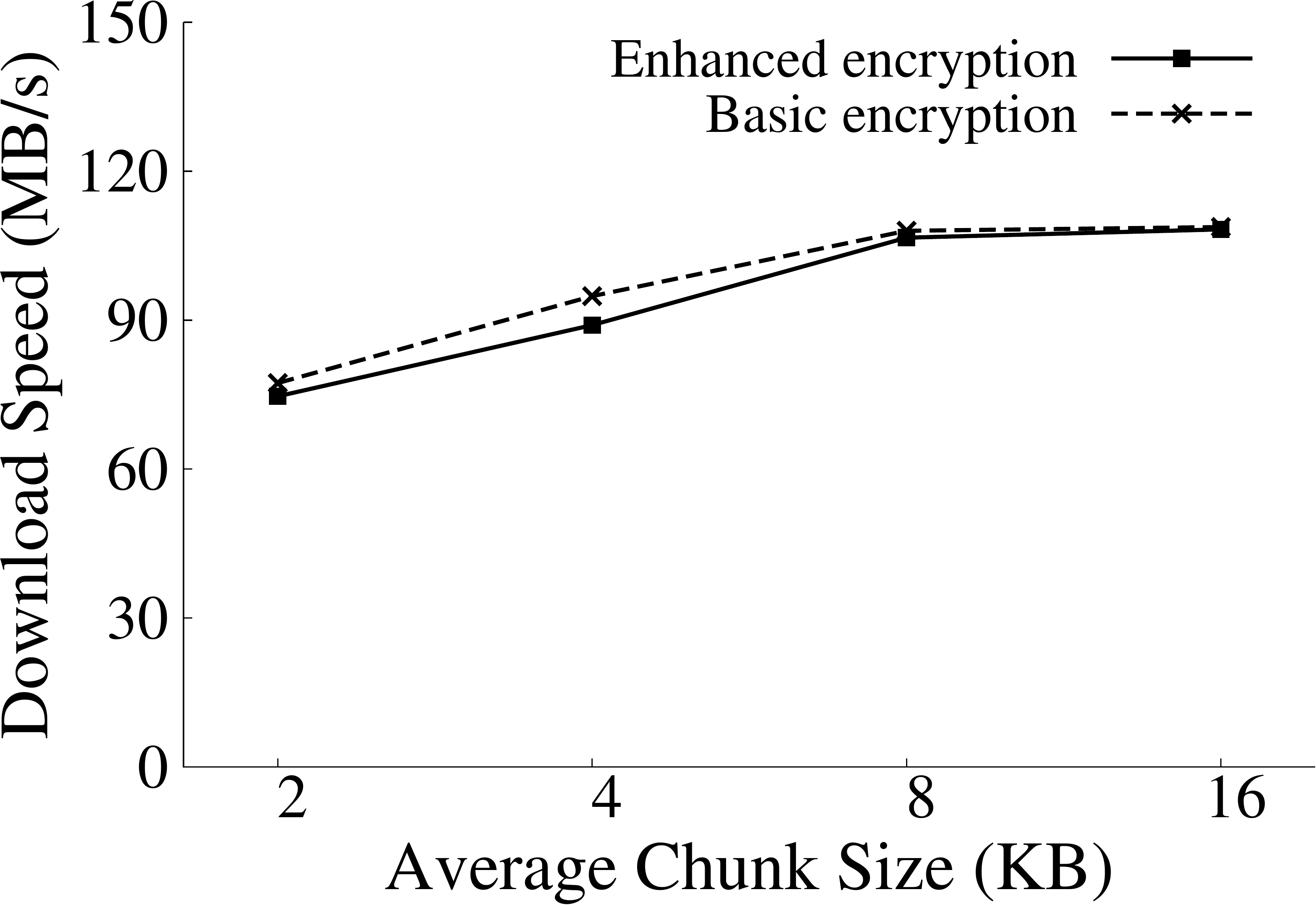}
	\label{fig:speed:down}
} \hspace{0.2in}
\subfigure[Aggregate upload speed]{
		\centering
	\includegraphics[width=2.5in]{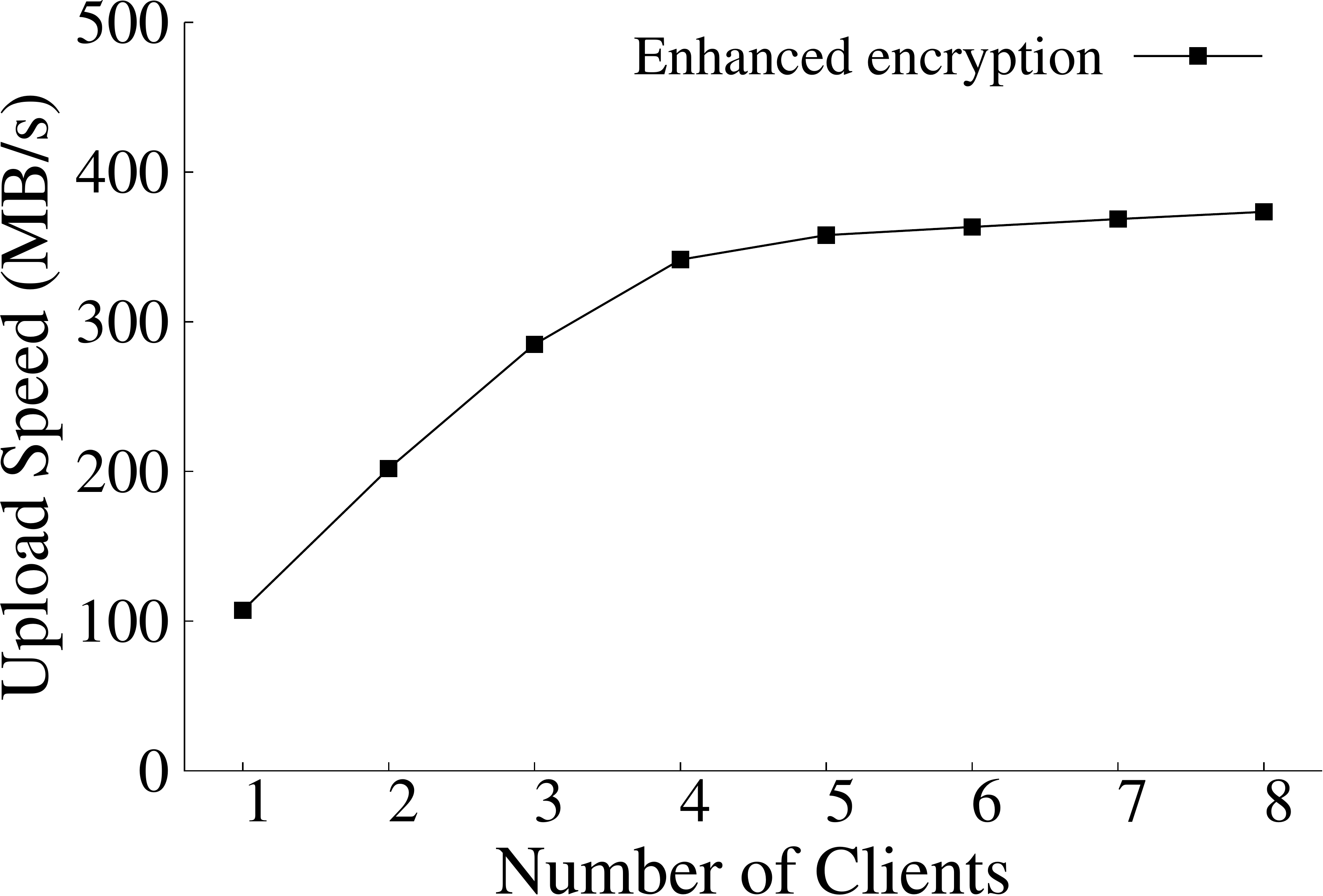}
	\label{fig:speed:multi}
}
\caption{Experiment A.3 (Upload and download performance).}
\end{figure}

Figure~\ref{fig:speed:up:chunk} shows the upload speeds under both encryption
schemes versus the average chunk size, in which we fix the average segment
size as 1MB. We see that the upload speeds increase with the average chunk
size, and become close to the effective network speed in our LAN testbed.  For
example, when the average chunk size is 16KB, the upload speeds for the
basic and enhanced schemes are 107.6MB/s and 106.9MB/s, respectively.  Both
encryption schemes have only minor performance differences.
%

Figure~\ref{fig:speed:up:segment} shows the upload speeds under both
encryption schemes versus the average segment size, in which we fix the
average chunk size as 8KB.  Similar to Figure~\ref{fig:speed:up:chunk}, the
upload speeds grow with the average segment size and are finally bounded by
the network speed. For example, when the average segment size is 1MB, the
upload speeds for the basic and enhanced schemes are 106.9MB/s and 106.4MB/s
respectively.

Figure~\ref{fig:speed:down} shows the download speeds under both encryption
schemes versus the average chunk size.  When the average chunk size goes
beyond 8KB, the download speeds of both encryption schemes (e.g., 108.0MB/s
for basic encryption and 106.6MB/s for enhanced encryption) approximate the
effective network speed.

We also consider the case with multiple REED clients.  We vary the number
of clients from one to eight, and each client runs on a different machine.
Here, we focus on the aggregate upload performance under the enhanced
encryption scheme, such that each client uploads a 2GB file of unique data
simultaneously.  We measure the {\em aggregate upload speed}, defined as the
ratio of the total amount of file data (i.e., 2GB times the number of clients)
to the total time when all uploads are finished.  Figure~\ref{fig:speed:multi}
shows the aggregate upload speed versus the number of clients, in which we fix
the average chunk size as 8KB and average segment size as 1MB.  We see that
the speed increases with the number of clients, and is finally bounded by the
network bandwidth. When there are eight clients, the aggregate upload speed
reaches 373.3MB/s. 

\paragraph{Experiment A.4 (Rekeying performance):} We measure the rekeying
performance in both lazy and active revocation schemes. Recall that the
rekeying operation of REED requires a CP-ABE decryption with the original
policy and another CP-ABE encryption with a new policy. REED treats each
policy as an access tree with an OR gate connecting all the authorized user
identifiers (see Section~\ref{subsec:policy}).  This implies that the CP-ABE
decryption time is constant \cite{bethencourt07}, while its encryption time
grows with the number of authorized users in the new policy.  
Thus, we focus on evaluating the impact of three parameters in the rekeying
operation: (i) {\em total number of users}, i.e., the number of authorized
users in the original policy; (ii) {\em revocation ratio}, the percentage of
the number of users to be revoked and removed from the access tree; and (iii)
{\em file size}, the size of the rekeyed file.  We measure the 
{\em rekeying delay}, defined as the total time of performing all rekeying
steps including: downloading and decrypting a key state, deriving a new key
state, encrypting and uploading the new key state, and re-encrypting the stub
file (for active revocation only). 

\begin{figure*}[t]
		\centering
		\subfigure[Varying the total number of users]{
			\includegraphics[width=2.5in]{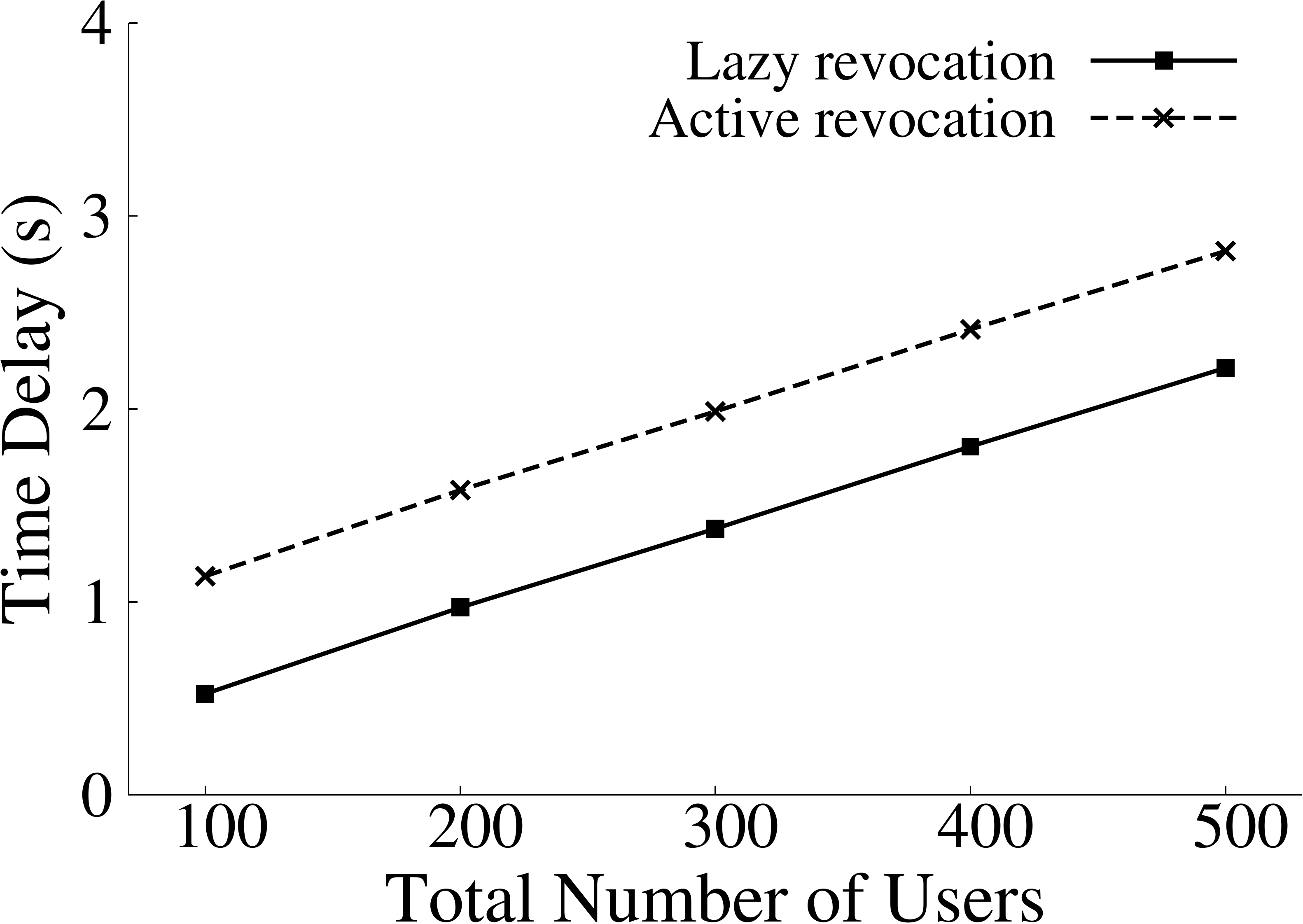}
			\label{fig:rekey:users}
		}\hspace{0.2in}
		\subfigure[Varying the revocation ratio]{
			\includegraphics[width=2.5in]{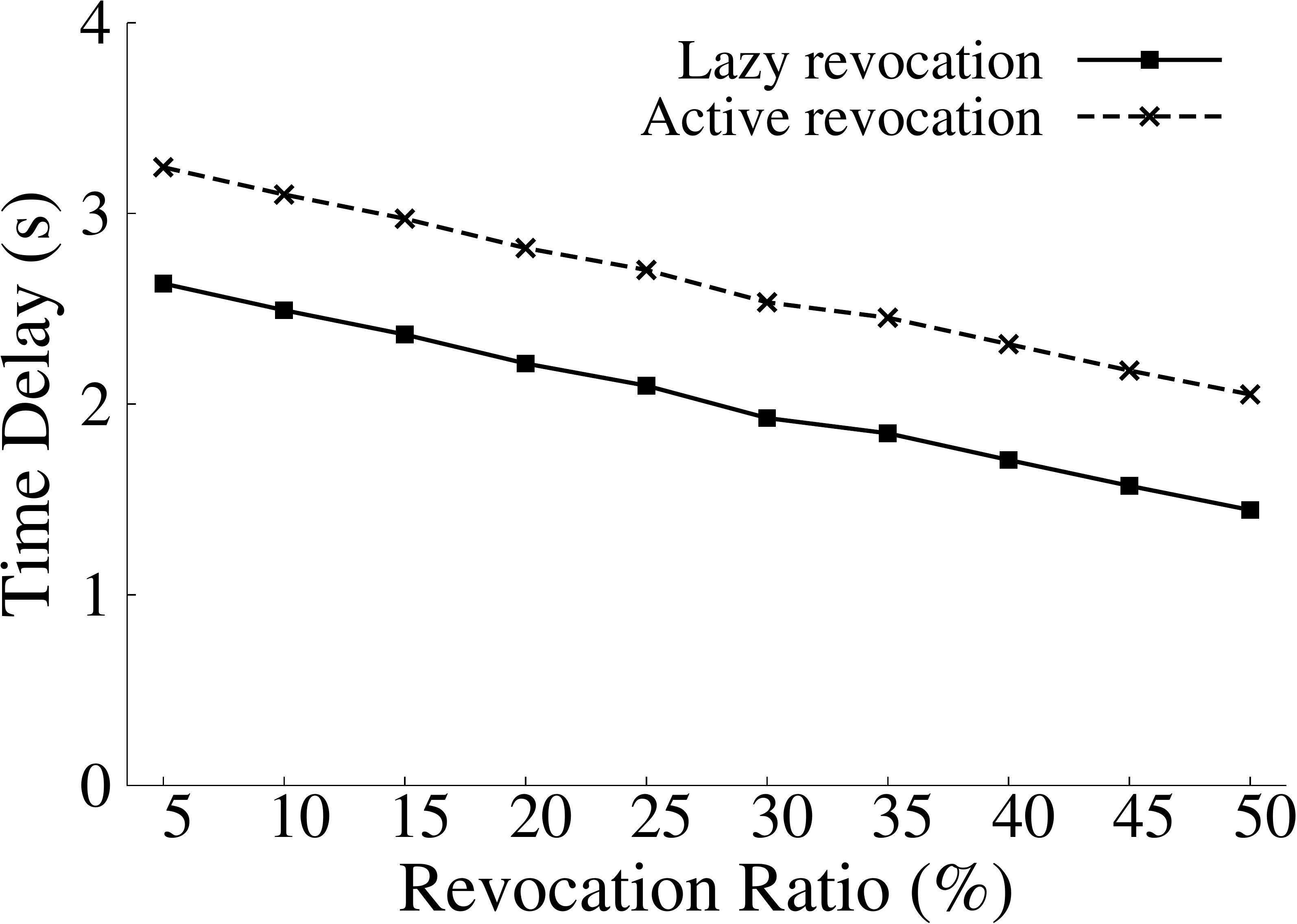}
		    \label{fig:rekey:ratio}
		}
		\subfigure[Varying the file size]{
			\includegraphics[width=2.5in]{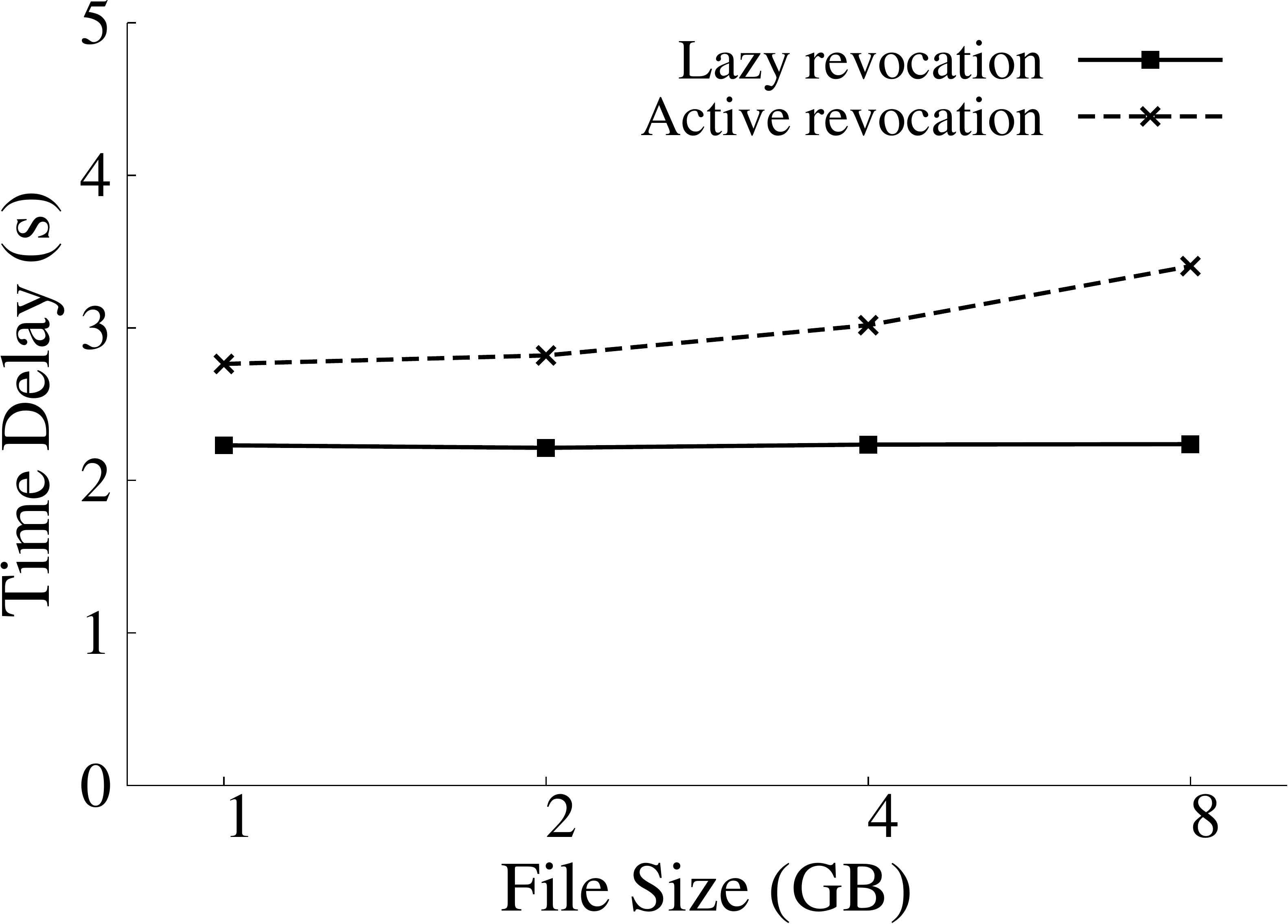}
			\label{fig:rekey:size}
		}
		\caption{Experiment~A.4 (Rekeying performance).}
\end{figure*}

Figure~\ref{fig:rekey:users} shows the rekeying delay versus the total
number of users, while we fix the rekeyed file size at 2GB and the revocation
ratio at 20\%.  The rekeying delays of both revocation schemes increase with
the total number of users, mainly because the CP-ABE encryption overhead
increases with a larger access tree.  Nevertheless, the rekeying delays are
within three seconds in both revocation schemes.  In particular, lazy
revocation is faster than active revocation by about 0.6s, as it defers
re-encryption process to the next file update. 

Figure~\ref{fig:rekey:ratio} shows the rekeying delay versus the revocation
ratio, while we fix the rekeyed file size at 2GB and the total number of users
at 500.  With a larger revocation ratio, the new policy has fewer authorized
users, thereby reducing the revocation time.  When the revocation ratio is
50\%, the rekeying delays of the lazy and active revocation schemes are 1.44s
and 2s, respectively. 

Figure~\ref{fig:rekey:size} shows the rekeying delay versus the size of the
rekeyed file, while we fix the total number of users at 500 and the revocation
ratio at 20\%.  The rekeyed file size has no impact on lazy revocation, in
which the rekeying delay is kept at 2.25s.  For active revocation, as the file
size increases, it spends more time for transferring and re-encrypting the
stub file.  Thus, the rekeying delay increases, for example, to 3.4s for an 8GB
file.  Nevertheless, if we compare the rekeying delay of active revocation
with the time of transferring a whole file in the network (e.g., at least 64s
in a 1Gb/s network), the rekeying delay is insignificant.  Thus, the rekeying
operation in REED is lightweight in general. 

\subsection{Real-world Data}
\label{subsec:real}

We now consider two real-world datasets to drive our evaluations.

\begin{itemize}
\item 
{\bf FSL:} This dataset is collected by the File systems and Storage Lab (FSL)
at Stony Brook University \cite{FSL14,sun16}.  The original FSL dataset contains
daily backups of the home directories of various users in a shared file
system.  We focus on the {\tt Fslhomes} dataset in 2013, which comprises 147
daily snapshots from January 22 to June 17, 2013.  Each snapshot represents a
daily backup, represented by a collection of 48-bit fingerprints of
variable-size chunks with an average 8KB chunk size. The dataset we consider
accounts for a total of 56.20TB of pre-deduplicated data. 
\item 
{\bf VM:} This dataset consists of virtual machine (VM) image snapshots and is
collected by ourselves.  We have 156 VMs for students enrolling in a
university programming course in Spring 2014.  We take 26 full image daily
snapshots for each VM spanning over three months.  Each image snapshot is of
size 10GB, and the complete dataset contains 39.61TB of data.  Each daily
snapshot is represented in SHA-1 fingerprints on 4KB fixed-size chunks.  We
remove all zero-filled chunks that are known to dominate in VM images
\cite{jin09}, and the size reduces to 18.24TB.  A subset of the same dataset
is also used in the prior work \cite{li15}.
\end{itemize}

In our evaluation, we construct variable-size segments with the average
segment size 1MB by grouping the chunks specified in the datasets (i.e., the
variable-size chunks in FSL and the fixed-size chunks in VM), based on
variable-size segmentation \cite{lillibridge09} described in
Section~\ref{sec:similarity}.



\paragraph{Experiment~B.1 (Storage overhead):}  We first measure the storage
overhead due to REED.  Our goal is to show that REED still maintains storage
efficiency via deduplication, even though it can only deduplicate part of a
chunk (i.e., trimmed package).  We define three types of data: (i) 
{\em logical data}, the original data before any encryption or deduplication;
(ii) {\em stub data}, the encrypted stub files being stored; (iii) 
{\em physical data}, the trimmed packages being stored after deduplication.
We aggregate the data from all users and measure the total size of each data
type.   

\begin{figure}[t]
\centering
\subfigure[Before and after deduplication (FSL)]{
    \includegraphics[width=2.5in]{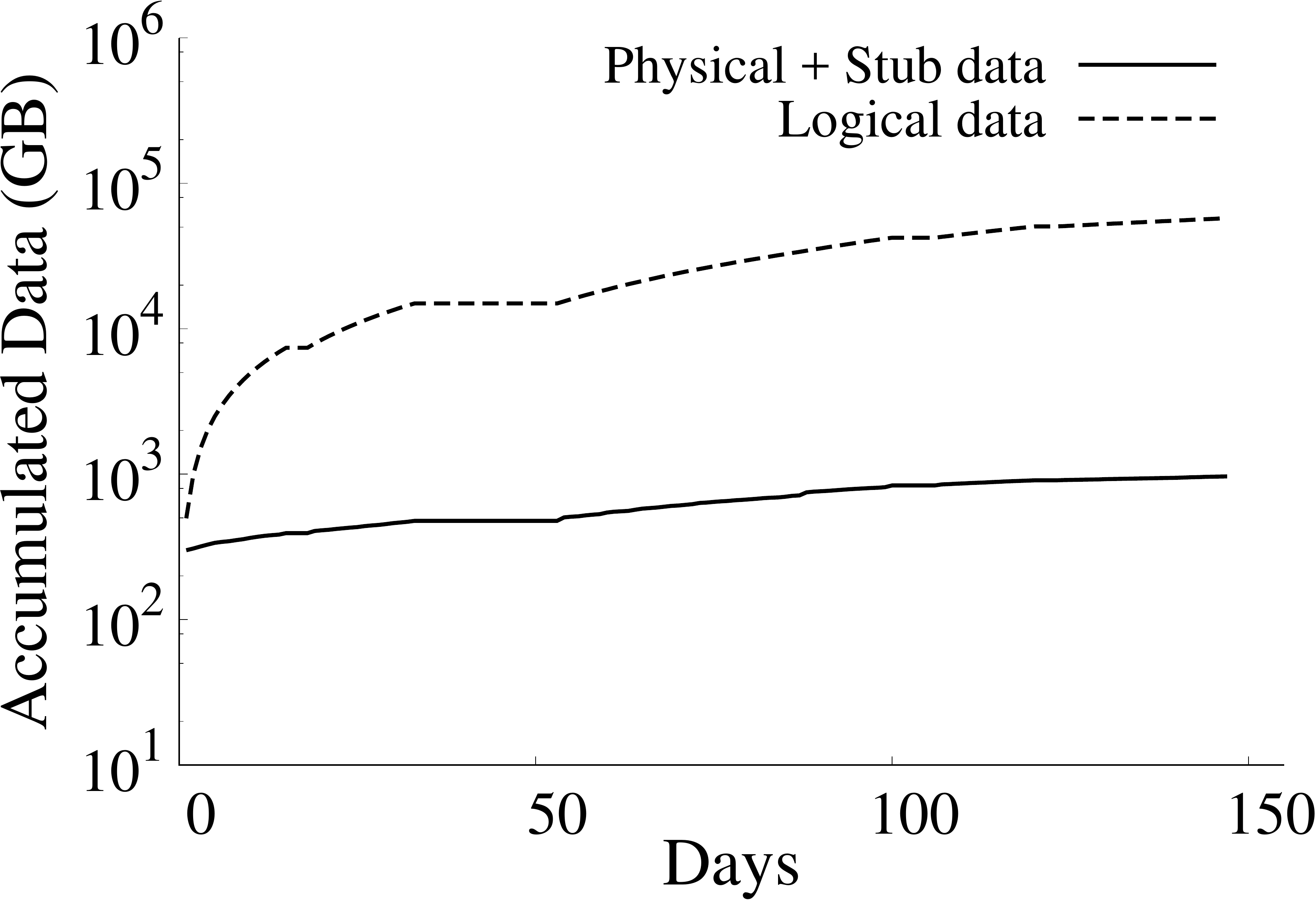}
	\label{fig:fsl:logic}
}\hspace{0.2in}
\subfigure[Physical and stub data (FSL)]{
    \includegraphics[width=2.5in]{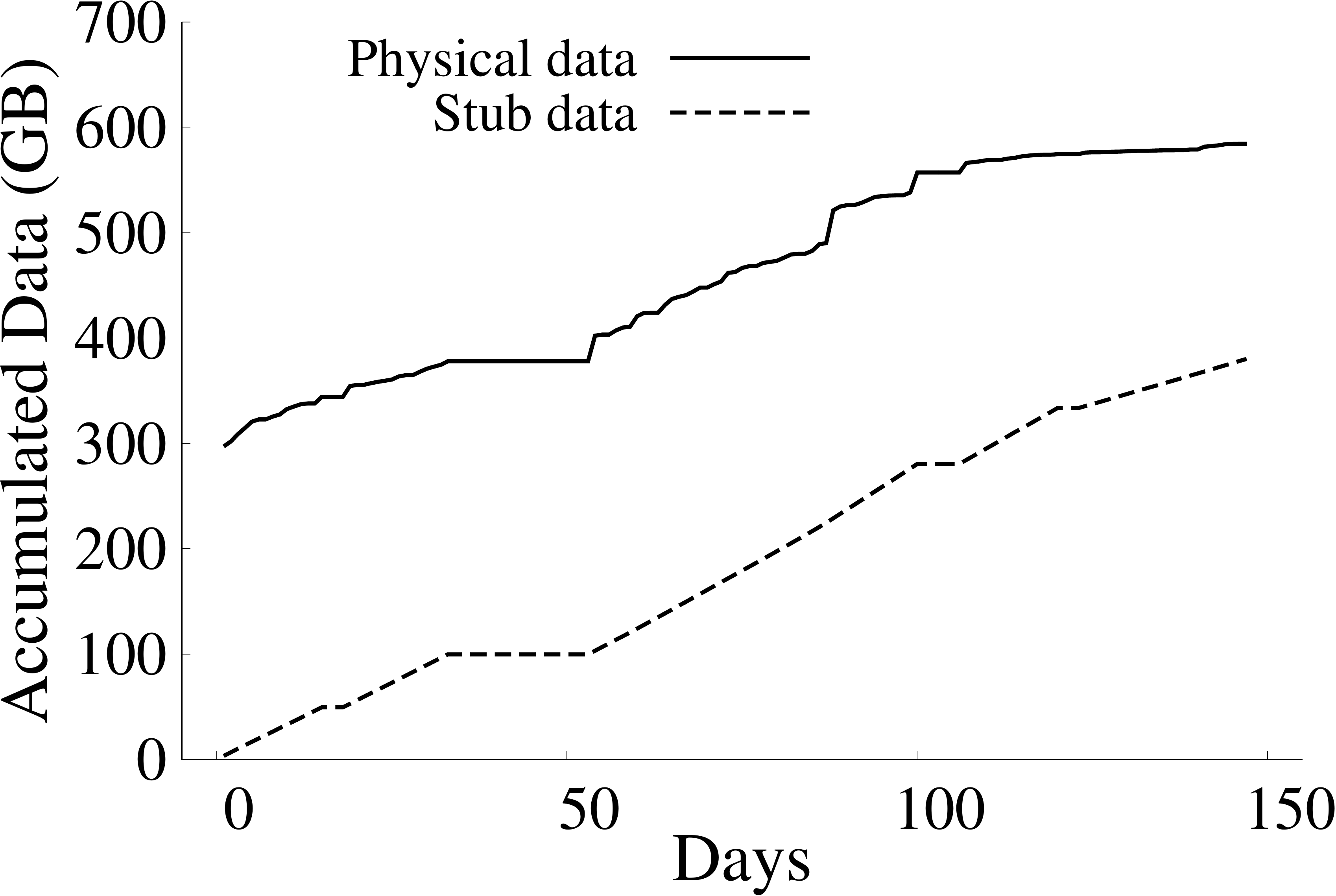}
	\label{fig:fsl:dedup}
}
\subfigure[Before and after deduplication (VM)]{
    \includegraphics[width=2.5in]{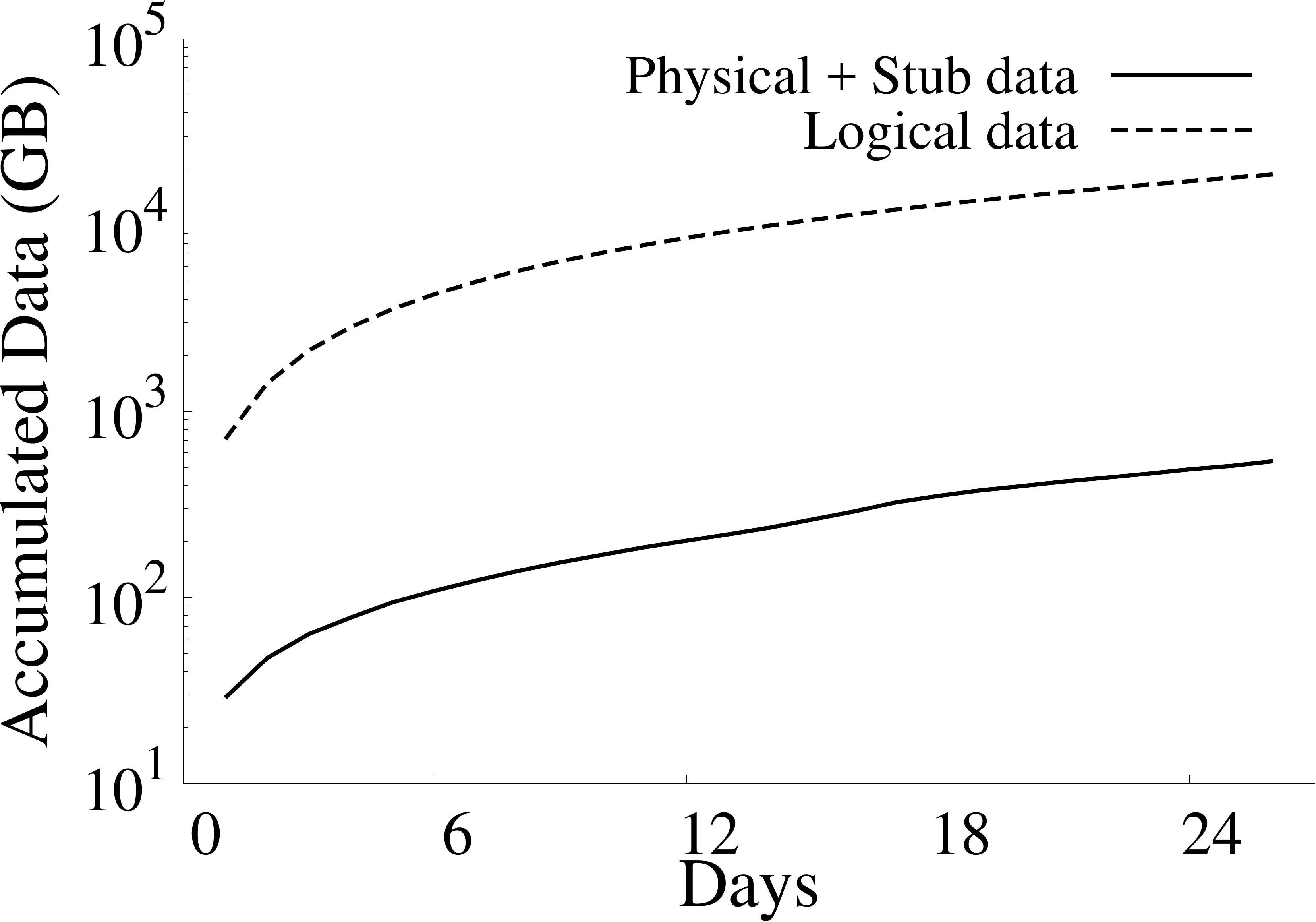}
	\label{fig:course:logic}
}\hspace{0.2in}
\subfigure[Physical and stub data (VM)]{
    \includegraphics[width=2.5in]{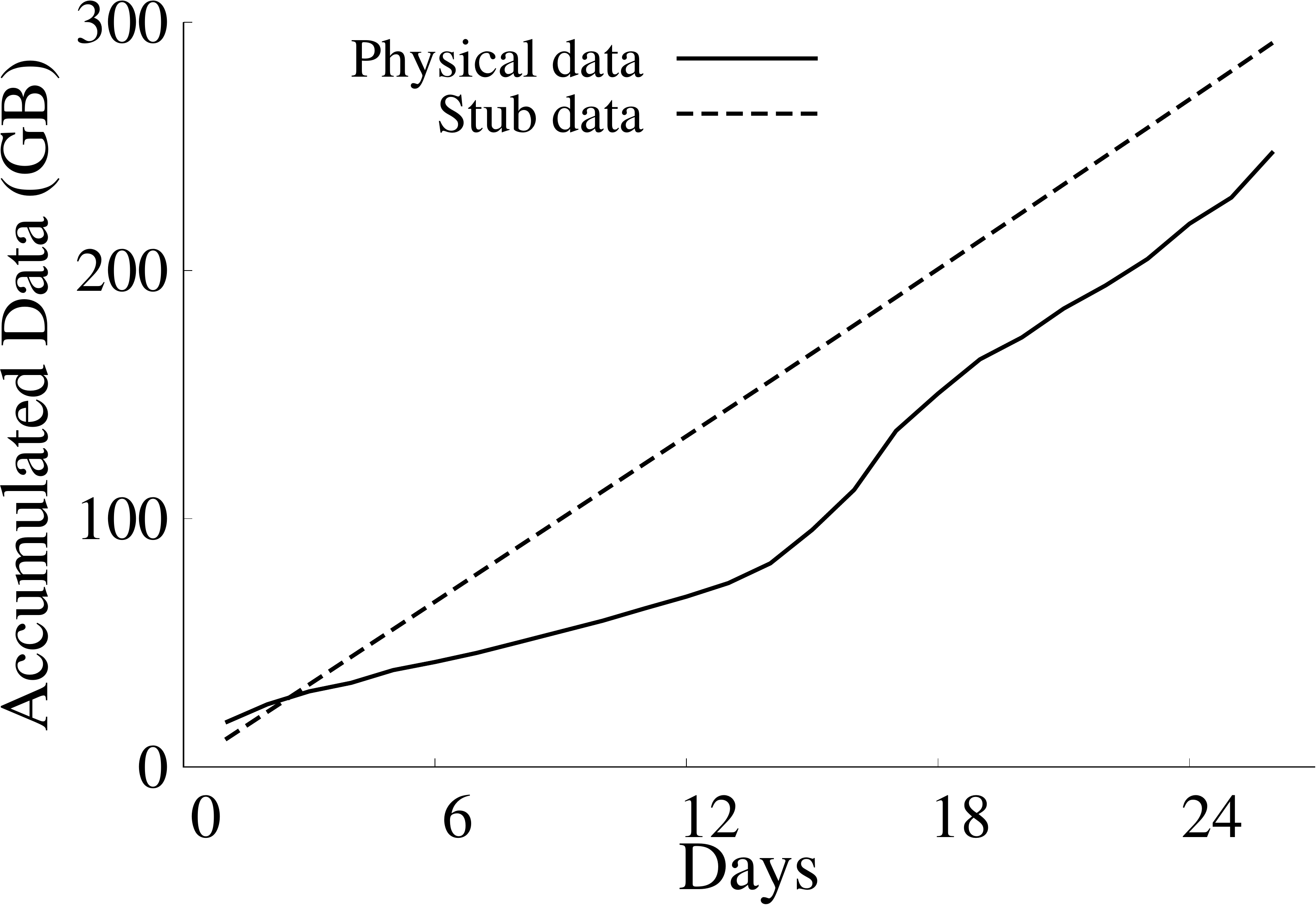}
	\label{fig:course:stub}
}
\caption{Experiment~B.1 (Storage overhead).}
\end{figure}

Figure~\ref{fig:fsl:logic} shows the cumulative data sizes over the number
of days of storing FSL daily backups of all users.  Each FSL daily backup
contains 290-680GB of logical data for all users, yet the physical and stub
data that REED actually stores after deduplication accounts for only 6.56GB
per day on average. After 147 days, there is a total of 57,548GB of logical
data, and REED generates only 964.4GB of physical and stub data after
deduplication.  It achieves a total saving of 98.3\%.  This shows that we
still maintain high storage efficiency through deduplication.  



Figure~\ref{fig:fsl:dedup} compares the cumulative sizes of physical and
stub data after deduplication. The cumulative size of stub data increases over
days. After 147 days,  there are 584.3GB of physical data due to the unique
trimmed packages. There is also 380.1GB of stub data.  Note that the stub data
cannot be deduplicated as it is encrypted by a renewable file key.
Nevertheless, deduplication effectively reduces the overall storage space
according to Figure~\ref{fig:fsl:logic}.

We now switch to the VM dataset.  Figure~\ref{fig:course:logic} compares the
size of logical data with the sizes of physical and stub data.  After 26 daily
backups, we have a total of 18,681GB of logical data.  Deduplication reduces
the space to 539.8GB for both physical data and stub data.  The storage saving
is 97.1\%.  Figure~\ref{fig:course:stub} presents a breakdown.  We observe
that the size of stub data grows linearly with the number of daily backups.
The reason is that the stub data size depends on the number of logical chunks,
yet each VM daily backup has a similar number of logical chunks (excluding the
zero-filled chunks). After 26 days, REED stores 247.9GB of physical data and
291.9GB of stub data.  The findings are similar to those for the FSL dataset. 




We further compare the storage overhead of our similarity-based approach with
that of the original {\em chunk-based approach}, which performs
deduplication at the granularity of chunks  (8KB for FSL and 4KB for VM).
Table~\ref{tab:comparison} shows the sizes of the physical and stub data, 
as well as the storage savings over the original size of logical data.  Our
similarity-based approach mitigates the MLE key generation overhead of the
chunk-based approach, while incurring 35.3\% and 45.8\% more size of physical
data.  Note that deduplication does not change the total number of logical
chunks, so both approaches have the same size of stub data for each dataset.
Nevertheless, the similarity-based approach still achieves almost identical
storage savings for both datasets as the chunk-based approach.

\begin{table}[!t]
\caption{Sizes of physical and stub data using different deduplication
approaches.}
\label{tab:comparison}
\renewcommand{\arraystretch}{1.2}
\vspace{-3pt}
\centering
\begin{small}
\begin{tabular}{|c|c|c|c|}
\hline
\multicolumn{2}{|c|}{\bf Data} & {\bf Chunk-based} & {\bf Similarity-based} \\
\hline
\hline
\multirow{2}{*}{FSL} & Physical  & 431.9GB  & 584.3GB \\
\cline{2-4}
			  & Stub & \multicolumn{2}{|c|}{380.1GB}\\
\hline
\multicolumn{2}{|c|}{\bf Storage saving} & 98.6\% & 98.3\% \\
\hline
\hline
\multirow{2}{*}{VM} & Physical & 170.0GB & 247.9GB \\
\cline{2-4}
			  & Stub & \multicolumn{2}{|c|}{291.9GB}\\
\hline
\multicolumn{2}{|c|}{\bf Storage saving} & 97.5\% & 97.1\%\\
\hline
\end{tabular}
\end{small}
\end{table}

REED focuses on maintaining high storage savings for logical data via
deduplication, yet we observe that stub data becomes dominant in physical
storage as more backups are stored (or more generally, for workloads with high
deduplication savings).  To mitigate the storage overhead of the stub data,
one option is to increase the chunk size; in fact, it has been shown that a
larger chunk size may achieve higher effective storage savings by reducing
metadata overhead \cite{sun16}.  We pose this issue as future work.

\paragraph{Experiment~B.2 (Trace-driven upload and download performance):} We
evaluate upload and download speeds of a single REED client using both
real-world datasets, as opposed to synthetic dataset in Experiment~A.3. 
Since both FSL and VM datasets only include chunk fingerprints and chunk
sizes, we reconstruct a chunk by repeatedly writing its fingerprint to a spare
chunk until reaching the specified chunk size; this ensures that the same
(resp. distinct) fingerprint returns the same (resp. distinct) chunk.  The
reconstructed chunk is treated as the output of chunking module of the REED
client. Thus, we do not include the chunking time in this experiment. 

The client uploads all daily backups (on behalf of all users), followed by
downloading them.  Due to the large dataset, we only run part of the dataset to
reduce the evaluation time. Specifically, for the FSL dataset, we choose seven
consecutive daily backups for nine users, totaling 3.64TB of data before
deduplication; for the VM dataset, we choose four daily backups for all users,
totaling 2.78TB of data before deduplication.  We use the same setting as in
Experiment~A.3, and use the enhanced encryption scheme.  

\begin{figure}[t]
\centering
\subfigure[FSL]{
\includegraphics[width=3.75in]{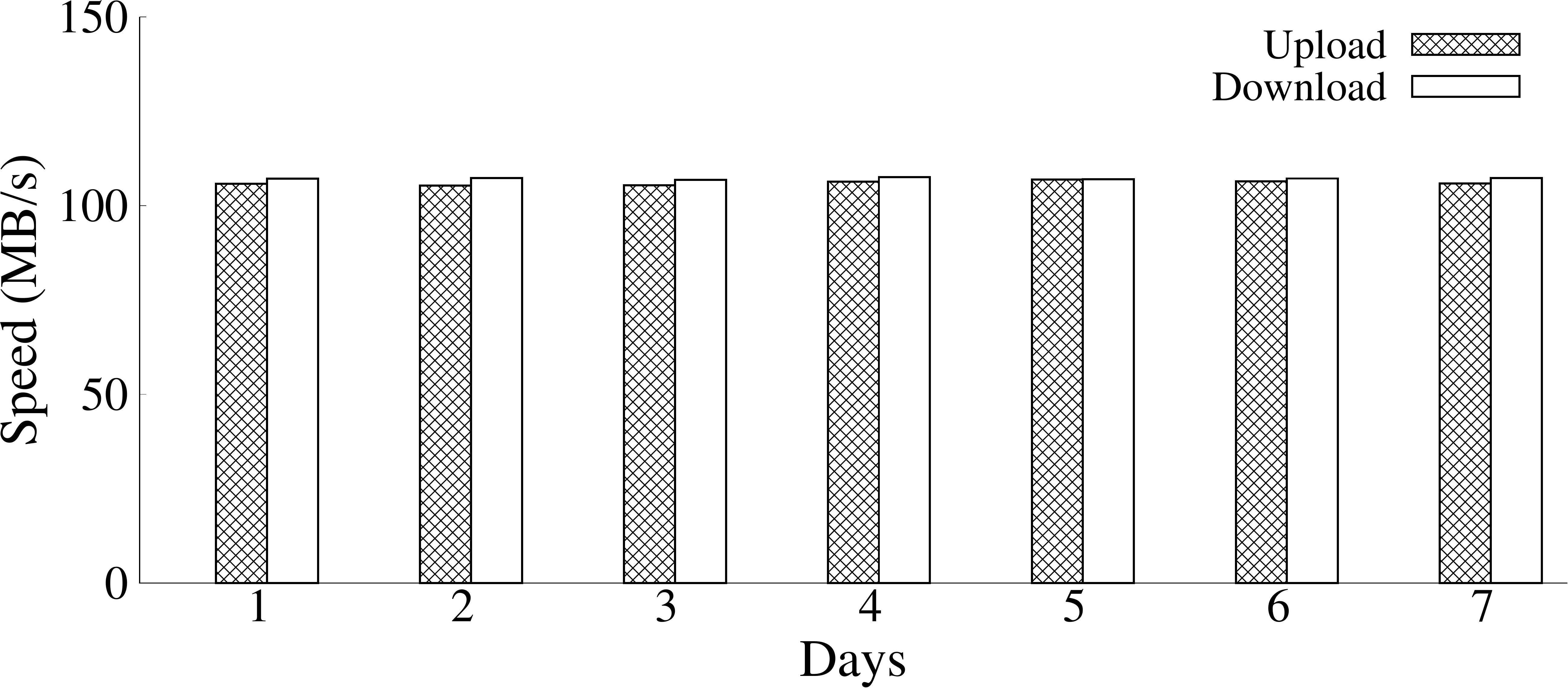}
}\hspace{0.2in}
\subfigure[VM]{
\includegraphics[width=2.5in]{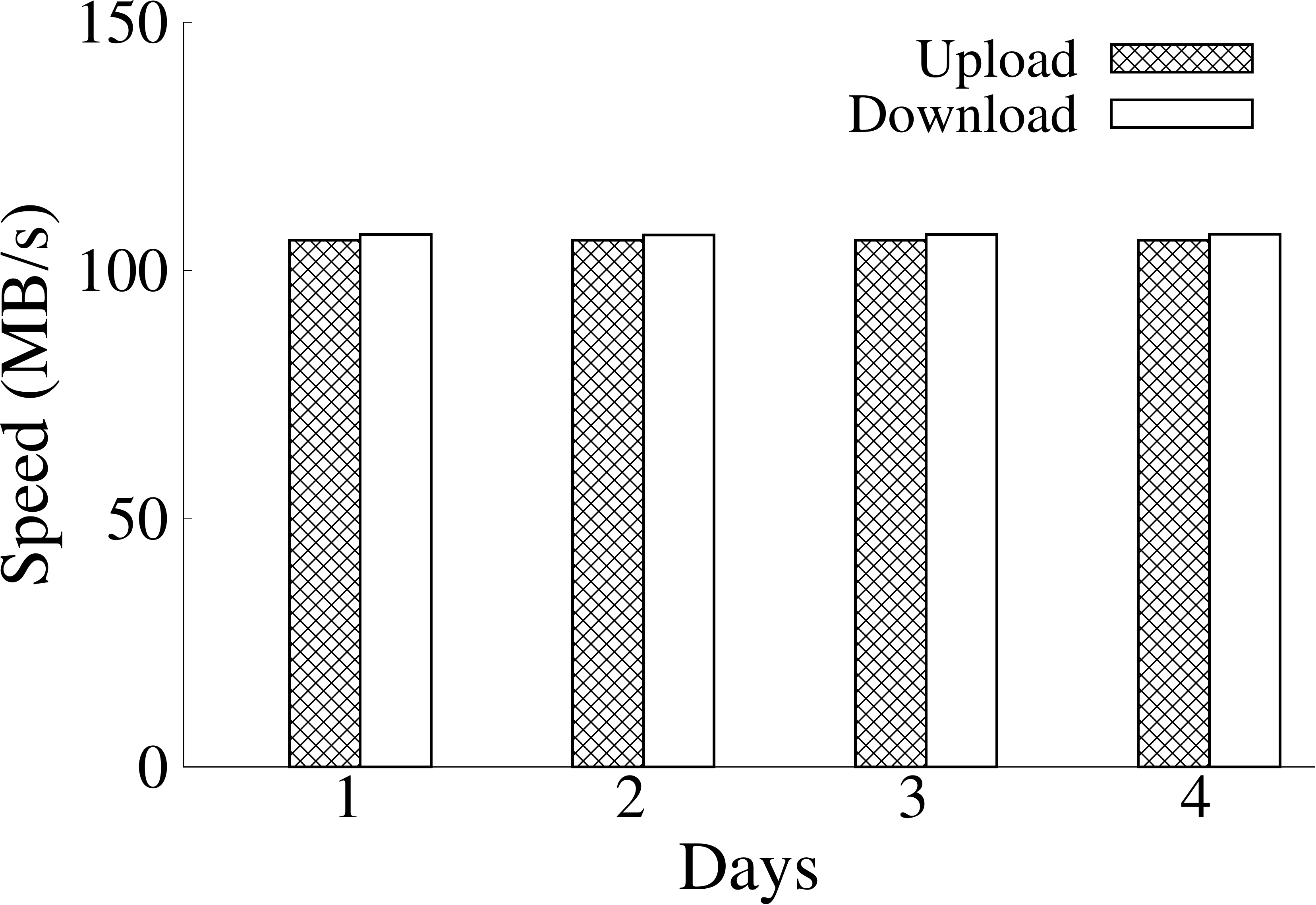}
}
\caption{Experiment~B.2 (Trace-driven upload and download performance).}
\label{fig:trace:tran}
\end{figure}

Figure~\ref{fig:trace:tran} shows the upload and download speeds of REED over
days.  Both the upload and download speeds of all days are almost
network-bound (around 105MB/s for both datasets) due to our segment-level MLE
key generation.  We highlight that for our original implementation in the
conference paper \cite{li16} the upload speed of the first day is as low as
13.1MB/s, since it lacks cached MLE keys and has to request MLE keys for each
chunk from the key manager.  Our similarity-based MLE key generation does not
have this limitation.

\section{Related Work}
\label{sec:related}

\paragraph{Encrypted deduplication storage:} Section~\ref{sec:background}
reviews MLE \cite{bellare13a} and DupLESS \cite{bellare13b},
which address the theoretical and applied aspects of encrypted deduplication
storage, respectively.  Bellare {\em et al.} \cite{bellare13a} propose a
theoretical framework of MLE, and provide formal definitions of privacy and
tag consistency.  The follow-up studies \cite{bellare15,abadi13} further
examine message correlation and parameter dependency of MLE.  

On the applied side, convergent encryption (CE) \cite{douceur02} has been
implemented and experimented in various storage systems (e.g.,
\cite{anderson10,wilcox-ohearn08,storer08,adya02,cox02,shah15}).
DupLESS \cite{bellare13b} implements server-aided MLE. Duan \cite{duan14}
improves the robustness of key management in DupLESS via threshold signature
\cite{shoup00}. 
Zheng {\em et al.} \cite{zheng15} propose a layer-level strategy specifically
for video deduplication.  Liu {\em et al.} \cite{liu15} propose a
password-authenticated key exchange protocol for MLE key generation.  ClearBox
\cite{armknecht15} enables clients to verify the effective storage
space that their data occupies after deduplication.  SecDep \cite{zhou15}
leverages cross-user file-level deduplication on the client side to mitigate
the key generation overhead, but it is susceptible to side channel attacks, in
which a malicious user can infer the existence of files through the
deduplication pattern \cite{harnik10,halevi11,li15}.  CDStore \cite{li15}
realizes CE in existing secret sharing algorithms by replacing the embedded
random seed with a message-derived hash to construct shares.  REED focuses on
the applied aspect, and complements the above designs by enabling rekeying in
encrypted deduplication storage. 

\paragraph{Rekeying:} 
Abdalla {\em et al.} \cite{abdalla00}
rigorously analyze key-derivation methods, in which a sequence of subkeys is
derived from a shared master key so as to extend the lifetime of the master
key for secure communication.  Follow-up studies examine key derivation (in
either key rotation or key regression) in content distribution networks
\cite{fu06,backes06,kallahall02} and cloud storage \cite{puttaswamy11}.  
A recent work \cite{watanabe13} examines ciphertext re-encryption using an 
approach similar to REED, in that it performs AONT on files and updates a
small piece from the AONT package, yet it does not consider deduplication and
has no prototype that demonstrates the applicability.  REED differs from
the above approaches by addressing the rekeying problem in encrypted
deduplication storage.  REED also uses the key regression scheme \cite{fu06}
in key derivation to enable lazy revocation. 

REED is related to secure deletion (see detailed surveys
\cite{diesburg10,reardon13}), which ensures that securely deleted data is
permanently inaccessible by anyone.
Secure deletion can be achieved through cryptographic deletion
(e.g., \cite{boneh96,peterson05}), which securely erases keys in
order to make encrypted data unrecoverable.  REED builds on the AONT-based
cryptographic deletion \cite{peterson05} and preserves deduplication
effectiveness. It further allows efficient dynamic access control.

\paragraph{Access control:}  Cryptographic primitives have been
proposed for enabling access control on encrypted storage, such as broadcast
encryption \cite{boneh05}, proxy re-encryption \cite{ateniese06}, and ABE
\cite{goyal06}.  REED builds on CP-ABE \cite{bethencourt07} to implement
fine-grained access control for encrypted deduplication storage.  

\paragraph{Exploiting workload characteristics:} Some studies address
deduplication performance by exploiting workload characteristics, including
chunk locality \cite{zhu08,kruus10}, similarity
\cite{lillibridge09,bhagwat09,dong11,fu12} and a combination of both
\cite{xia11}.  REED is motivated from a security perspective, and uses the
similarity-based approach in Extreme Binning \cite{bhagwat09} to mitigate MLE
key generation overhead.  



\section{Conclusion}
\label{sec:conclusions}

We present REED, an encrypted deduplication storage system that aims for
secure and lightweight rekeying.  The core rekeying design of REED is to renew
a key of a deterministic all-or-nothing-transform (AONT) package.  We propose
two encryption schemes for REED: the basic scheme has higher encryption
performance, while the enhanced scheme is resilient against key leakage.  We
extend REED with dynamic access control by integrating both CP-ABE and key
regression primitives.  We show the confidentiality and integrity properties
of REED under our security definitions. Furthermore, we propose a
similarity-based approach to mitigate MLE key generation overhead of REED.
We finally implement a REED prototype, and conduct trace-driven evaluation in
a LAN testbed to demonstrate its performance and storage efficiency. 
In future work, we plan to address the open issues of our current REED design
(see Section~\ref{subsec:discussion}), investigate how we mitigate the storage
overhead of stub data (see Section~\ref{subsec:real}), and evaluate how REED
performs for other storage workloads.

\bibliographystyle{plain}
\bibliography{references}

\begin{thebibliography}{10}

\bibitem{FSL14}
{FSL} traces and snapshots public archive.
\newblock http://tracer.filesystems.org/, 2014.

\bibitem{abadi13}
Mart{\'\i}n Abadi, Dan Boneh, Ilya Mironov, Ananth Raghunathan, and Gil Segev.
\newblock Message-locked encryption for lock-dependent messages.
\newblock In {\em Proc. of CRYPTO}, 2013.

\bibitem{abdalla00}
Michel Abdalla and Mihir Bellare.
\newblock Increasing the lifetime of a key: A comparative analysis of the
  security of re-keying techniques.
\newblock In {\em Proc. of ASIACRYPT}, 2000.

\bibitem{adya02}
Atul Adya, William~J. Bolosky, Miguel Castro, Gerald Cermak, Ronnie Chaiken,
  John~R. Douceur, Jon Howell, Jacob~R. Lorch, Marvin Theimer, and Roger~P.
  Wattenhofer.
\newblock Farsite: Federated, available, and reliable storage for an
  incompletely trusted environment.
\newblock In {\em Proc. of USENIX OSDI}, 2002.

\bibitem{amazon14}
Amazon.
\newblock Architecting for genomic data security and compliance in {AWS}, 2014.

\bibitem{anderson10}
Paul Anderson and Le~Zhang.
\newblock Fast and secure laptop backups with encrypted de-duplication.
\newblock In {\em Proc. of USENIX LISA}, 2010.

\bibitem{armknecht15}
Frederik Armknecht, Jens-Matthias Bohli, Ghassan~O. Karame, and Franck Youssef.
\newblock Transparent data deduplication in the cloud.
\newblock In {\em Proc. of ACM CCS}, 2015.

\bibitem{ateniese07}
Giuseppe Ateniese, Randal Burns, Reza Curtmola, Joseph Herring, Lea Kissner,
  Zachary Peterson, and Dawn Song.
\newblock Provable data possession at untrusted stores.
\newblock In {\em Proc. of ACM CCS}, 2007.

\bibitem{ateniese06}
Giuseppe Ateniese, Kevin Fu, Matthew Green, and Susan Hohenberger.
\newblock Improved proxy re-encryption schemes with applications to secure
  distributed storage.
\newblock {\em ACM Trans. Inf. Syst. Secur.}, 9(1):1--30, February 2006.

\bibitem{backes06}
Michael Backes, Christian Cachin, and Alina Oprea.
\newblock Secure key-updating for lazy revocation.
\newblock In {\em Proc. of ESORICS}, 2006.

\bibitem{barker12}
Elaine Barker, William Barker, William Burr, William Polk, and Miles Smid.
\newblock {NIST Special Publication 800-57} recommendation for key management.
\newblock Technical report, National Institute of Standards \& Technology,
  {July} 2012.

\bibitem{bellare15}
Mihir Bellare and Sriram Keelveedhi.
\newblock Interactive message-locked encryption and secure deduplication.
\newblock In {\em Proc. of PKC}, 2015.

\bibitem{bellare13b}
Mihir Bellare, Sriram Keelveedhi, and Thomas Ristenpart.
\newblock {DupLESS}: Server-aided encryption for deduplicated storage.
\newblock In {\em Proc. of USENIX Security}, 2013.

\bibitem{bellare13a}
Mihir Bellare, Sriram Keelveedhi, and Thomas Ristenpart.
\newblock Message-locked encryption and secure deduplication.
\newblock In {\em Proc. of EUROCRYPT}, 2013.

\bibitem{bethencourt07}
John Bethencourt, Amit Sahai, and Brent Waters.
\newblock Ciphertext-policy attribute-based encryption.
\newblock In {\em IEEE S\&P}, 2007.

\bibitem{bethencourt_abetoolkit}
John Bethencourt, Amit Sahai, and Brent Waters.
\newblock {CP-ABE} toolkit.
\newblock http://acsc.cs.utexas.edu/cpabe/, 2011.

\bibitem{bhagwat09}
Deepavali Bhagwat, Kave Eshghi, Darrell~D.E. Long, and Mark Lillibridge.
\newblock Extreme binning: Scalable, parallel deduplication for chunk-based
  file backup.
\newblock In {\em Proc. of IEEE MASCOTS}, 2009.

\bibitem{black06}
John Black.
\newblock Compare-by-hash: a reasoned analysis.
\newblock In {\em Proc. of USENIX ATC}, 2006.

\bibitem{boneh05}
Dan Boneh, Craig Gentry, and Brent Waters.
\newblock Collusion resistant broadcast encryption with short ciphertexts and
  private keys.
\newblock In {\em Proc. of CRYPTO}, 2005.

\bibitem{boneh96}
Dan Boneh and Richard Lipton.
\newblock A revocable backup system.
\newblock In {\em Proc. of USENIX Security}, 1996.

\bibitem{boneh01}
Dan Boneh, Ben Lynn, and Hovav Shacham.
\newblock Short signatures from the weil pairing.
\newblock In {\em Proc. of ASIACRYPT}, 2001.

\bibitem{broder97}
Andrei~Z. Broder.
\newblock On the resemblance and containment of documents.
\newblock In {\em Proc. of IEEE Compression and Complexity of Sequences}, 1997.

\bibitem{cox02}
Landon~P. Cox, Christopher~D. Murray, and Brian~D. Noble.
\newblock Pastiche: Making backup cheap and easy.
\newblock In {\em Proc. of USENIX OSDI}, 2002.

\bibitem{csaplar11}
Dick Csaplar.
\newblock Building business resillience through active archiving, 2011.

\bibitem{debian-security-advisory08}
{Debian Security Advisory}.
\newblock {DSA}-1571-1 openssl -- predictable random number generator.
\newblock https://www.debian.org/security/2008/dsa-1571, May 2008.

\bibitem{diesburg10}
Sarah~M. Diesburg and An-I~Andy Wang.
\newblock A survey of confidential data storage and deletion methods.
\newblock {\em ACM Comput. Surv.}, 43(1):2:1--2:37, December 2010.

\bibitem{dong11}
Wei Dong, Fred Douglis, Kai Li, and Hugo Patterson.
\newblock Tradeoffs in scalable data routing for deduplication clusters.
\newblock In {\em Proc. of USENIX FAST}, 2011.

\bibitem{douceur02}
John~R. Douceur, Atul Adya, William~J. Bolosky, Dan Simon, and Marvin Theimer.
\newblock Reclaiming space from duplicate files in a serverless distributed
  file system.
\newblock In {\em Proc. of IEEE ICDCS}, 2002.

\bibitem{duan14}
Yitao Duan.
\newblock Distributed key generation for encrypted deduplication: Achieving the
  strongest privacy.
\newblock In {\em Proc. of ACM CCSW}, 2014.

\bibitem{fu06}
Kevin Fu, Seny Kamara, and Tadayoshi Kohno.
\newblock Key regression: Enabling efficient key distribution for secure
  distributed storage.
\newblock In {\em Proc. of NDSS}, 2006.

\bibitem{fu12}
Yinjin Fu, Hong Jiang, and Nong Xiao.
\newblock A scalable inline cluster deduplication framework for big data
  protection.
\newblock In {\em Proc. of Middleware}, 2012.

\bibitem{goldwasser08}
Shafi Goldwasser and Mihir Bellare.
\newblock Lecture notes on cryptography.
\newblock https://cseweb.ucsd.edu/~mihir/papers/gb.html, {July} 2008.

\bibitem{googlegenomics}
Google.
\newblock Google genomics.
\newblock https://cloud.google.com/genomics/, 2016.

\bibitem{goyal06}
Vipul Goyal, Omkant Pandey, Amit Sahai, and Brent Waters.
\newblock Attribute-based encryption for fine-grained access control of
  encrypted data.
\newblock In {\em Proc. of ACM CCS}, 2006.

\bibitem{halevi11}
Shai Halevi, Danny Harnik, Benny Pinkas, and Alexandra Shulman-Peleg.
\newblock Proofs of ownership in remote storage systems.
\newblock In {\em Proc. of ACM CCS}, 2011.

\bibitem{harnik10}
Danny Harnik, Benny Pinkas, and Alexandra Shulman-Peleg.
\newblock Side channels in cloud services: Deduplication in cloud storage.
\newblock {\em IEEE Security \& Privacy}, 8(6):40--47, 2010.

\bibitem{jin09}
Keren Jin and Ethan~L. Miller.
\newblock The effectiveness of deduplication on virtual machine disk images.
\newblock In {\em Proc. of ACM SYSTOR}, 2009.

\bibitem{juels07}
Ari Juels and Burton~S. Kaliski, Jr.
\newblock {PORs}: Proofs of retrievability for large files.
\newblock In {\em Proc. of ACM CCS}, 2007.

\bibitem{kallahall02}
Mahesh Kallahall, Erik Riedel, Ram Swaminathan, Qian Wang, and Kevin Fu.
\newblock Plutus: Scalable secure file sharing on untrusted storage.
\newblock In {\em Proc. of USENIX FAST}, 2002.

\bibitem{kaminsky11}
Dan Kaminsky.
\newblock These are not the certs you're looking for.
\newblock http://dankaminsky.com/2011/08/31/notnotar/, Aug 2011.

\bibitem{kruus10}
Erik Kruus, Cristian Ungureanu, and Cezary Dubnicki.
\newblock Bimodal content defined chunking for backup streams.
\newblock In {\em Proc. of USENIX FAST}, 2010.

\bibitem{li16}
Jingwei Li, Chuan Qin, Patrick P.~C. Lee, and Jin Li.
\newblock Rekeying for encrypted deduplication storage.
\newblock In {\em IEEE/IFIP DSN}, 2016.

\bibitem{li15}
Mingqiang Li, Chuan Qin, and Patrick P.~C. Lee.
\newblock {CDStore}: Toward reliable, secure, and cost-efficient cloud storage
  via convergent dispersal.
\newblock In {\em Proc. of USENIX ATC}, 2015.

\bibitem{lillibridge09}
Mark Lillibridge, Kave Eshghi, Deepavali Bhagwat, Vinay Deolalikar, Greg
  Trezise, and Peter Camble.
\newblock Sparse indexing: Large scale, inline deduplication using sampling and
  locality.
\newblock In {\em Proc. of USENIX FAST}, 2009.

\bibitem{liu15}
Jian Liu, N.~Asokan, and Benny Pinkas.
\newblock Secure deduplication of encrypted data without additional independent
  servers.
\newblock Cryptology ePrint Archive: Report 2015/455, August 2015.

\bibitem{Cloudencryption13}
Linda Musthaler.
\newblock Cloud encryption: Control your own keys in a separate storage vault.
\newblock
  http://www.networkworld.com/article/2170564/cloud-computing/cloud-encryption-control-your-own-keys-in-a-separate-storage-vault.html,
  2013.

\bibitem{NIH15}
{National Institutes of Health}.
\newblock {NIH} security best practices for controlled-access data subject to
  the {NIH} genomic data sharing policy, 2015.

\bibitem{NetApp08}
NetApp.
\newblock Netapp deduplication helps duke institute for genome sciences and
  policy reduce storage requirements for genomic information by 83 percent.
\newblock
  http://www.netapp.com/us/company/news/press-releases/news-rel-20081008.aspx,
  2008.

\bibitem{openssl}
{OpenSSL}.
\newblock https://www.openssl.org, 2015.

\bibitem{peterson05}
Zachary N.~J. Peterson, Randal Burns, Joe Herring, Adam Stubblefield, and
  Aviel~D. Rubin.
\newblock Secure deletion for a versioning file system.
\newblock In {\em Proc. of USENIX FAST}, 2005.

\bibitem{puttaswamy11}
Krishna~PN Puttaswamy, Christopher Kruegel, and Ben~Y Zhao.
\newblock Silverline: toward data confidentiality in storage-intensive cloud
  applications.
\newblock In {\em Proc. of ACM SoCC}, 2011.

\bibitem{rabin81}
Michael~O. Rabin.
\newblock Fingerprinting by random polynomials.
\newblock Center for Research in Computing Technology, Harvard University.
  Tech. Report TR-CSE-03-01, 1981.

\bibitem{rahumed11}
A.~Rahumed, H.C.H. Chen, Yang Tang, P.~P.~C. Lee, and J.C.S. Lui.
\newblock A secure cloud backup system with assured deletion and version
  control.
\newblock In {\em Proc. of IEEE ICPPW}, Sept 2011.

\bibitem{reardon13}
Joel Reardon, David Basin, and Srdjan Capkun.
\newblock {SoK}: Secure data deletion.
\newblock In {\em Proc. of IEEE S\&P}, 2013.

\bibitem{resch11}
Jason~K. Resch and James~S. Plank.
\newblock {AONT-RS}: Blending security and performance in dispersed storage
  systems.
\newblock In {\em Proc. of USENIX FAST}, 2011.

\bibitem{rivest97}
Ronald~L. Rivest.
\newblock All-or-nothing encryption and the package transform.
\newblock In {\em Proc. of FSE}, 1997.

\bibitem{shah15}
Peter Shah and Won So.
\newblock Lamassu: Storage-efficient host-side encryption.
\newblock In {\em Proc. of USENIX ATC}, 2015.

\bibitem{shoup00}
Victor Shoup.
\newblock Practical threshold signatures.
\newblock In {\em Proc. of EUROCRYPT}, 2000.

\bibitem{sotirov08}
Alexander Sotirov, Marc Stevens, Jacob Appelbaum, Arjen Lenstra, David Molnar,
  Dag~Arne Osvik, and Benne de~Weger.
\newblock Md5 considered harmful today.
\newblock http://www.win.tue.nl/hashclash/rogue-ca/, Dec 2008.

\bibitem{stein10}
Lincoln~D Stein.
\newblock The case for cloud computing in genome informatics.
\newblock {\em Genome Biology}, 2010.

\bibitem{storer08}
Mark~W. Storer, Kevin Greenan, Darrell~D.E. Long, and Ethan~L. Miller.
\newblock Secure data deduplication.
\newblock In {\em Proc. of ACM StorageSS}, 2008.

\bibitem{sun16}
Zhu Sun, Geoff Kuenning, Sonam Mandal, Philip Shilane, Vasily Tarasov, Nong
  Xiao, and Erez Zadok.
\newblock A long-term user-centric analysis of deduplication patterns.
\newblock In {\em Proc. of IEEE MSST}, 2016.

\bibitem{us-computer-emergency-readiness-team14}
{U.S. Computer Emergency Readiness Team}.
\newblock {OpenSSL} `heartbleed' vulnerability ({CVE}-2014-0160).
\newblock https://www.us-cert.gov/ncas/alerts/TA14-098A, April 2014.

\bibitem{wallace12}
Grant Wallace, Fred Douglis, Hangwei Qian, Philip Shilane, Stephen Smaldone,
  Mark Chamness, and Windsor Hsu.
\newblock Characteristics of backup workloads in production systems.
\newblock In {\em Proc. of USENIX FAST}, 2012.

\bibitem{watanabe13}
Dai Watanabe and Masayuki Yoshino.
\newblock Key update mechanism for network storage of encrypted data.
\newblock In {\em Proc. of IEEE CloudCom}, 2013.

\bibitem{webster85}
A.~F. Webster and S.~E. Tavares.
\newblock On the design of {S}-boxes.
\newblock In {\em Proc. of CRYPTO}, 1985.

\bibitem{wilcox-ohearn08}
Zooko Wilcox-O'Hearn and Brian Warner.
\newblock Tahoe: The least-authority filesystem.
\newblock In {\em Proc. of ACM StorageSS}, 2008.

\bibitem{xia11}
Wen Xia, Hong Jiang, Dan Feng, and Yu~Hua.
\newblock {SiLo}: A similarity locality based near exact deduplication scheme
  with low ram overhead and high throughput.
\newblock In {\em Proc. of USENIX ATC}, 2011.

\bibitem{zheng15}
Yifeng Zheng, Xingliang Yuan, Xinyu Wang, Jinghua Jiang, Cong Wang, and Xiaolin
  Gui.
\newblock Enabling encrypted cloud media center with secure deduplication.
\newblock In {\em Proc. of ACM ASIACCS}, 2015.

\bibitem{zhou15}
Yukun Zhou, Dan Feng, Wen Xia, Min Fu, Fangting Huang, Yucheng Zhang, and
  Chunguang Li.
\newblock {SecDep}: A user-aware efficient fine-grained secure deduplication
  scheme with multi-level key management.
\newblock In {\em Proc. of IEEE MSST}, 2015.

\bibitem{zhu08}
Benjamin Zhu, Kai Li, and R~Hugo Patterson.
\newblock Avoiding the disk bottleneck in the data domain deduplication file
  system.
\newblock In {\em Proc. of USENIX FAST}, 2008.

\end{thebibliography}

\end{document}